\documentclass[preprint]{elsarticle}
\pdfoutput=1
\usepackage{lineno,hyperref}
\usepackage{textcomp}
\usepackage{graphicx}
\usepackage{siunitx}
\usepackage{subfig}
\usepackage{ifpdf}
\usepackage{xcolor}

\modulolinenumbers[5]

\journal{Nuclear Instruments and Methods in Physics Research Section A, }

\begin{document}

\begin{frontmatter}

\title{Energy Dependent Features of X-ray Signals in a GridPix Detector}


\author[mymainaddress]{C. Krieger}
\author[mymainaddress]{J. Kaminski\corref{mycorrespondingauthor}}
\cortext[mycorrespondingauthor]{Corresponding author}\ead{kaminski@physik.uni-bonn.de}
\author[CERN]{T. Vafeiadis}
\author[mymainaddress]{K. Desch}

\address[mymainaddress]{Physikalisches Institut, University of Bonn, Germany}
\address[CERN]{CERN, Geneva, Switzerland}

\begin{abstract}
We report on the calibration of an argon/isobutane (\SI{97.7}{\percent}/\SI{2.3}{\percent})-filled GridPix detector with soft X-rays (\SI{277}{\eV} to \SI{8}{\keV}) using the variable energy X-ray source of the CAST Detector Lab at CERN. We study the linearity and energy resolution of the detector using both the number of hit pixels and the total measured charge as energy measures. For the latter, the energy resolution $\sigma_E/E$ is better than \SI{10}{\percent} (\SI{20}{\percent}) for energies above \SI{2}{\keV} (\SI{0.5}{\keV}). Several characteristics of the recorded events are studied.
\end{abstract}


\end{frontmatter}


\section{Introduction and Description of Setup}\label{sec_intro}
Gaseous X-ray detectors with a high granularity readout have advantages over other detector types, such as silicon detectors, if additional
information based on the event shape is required to suppress non-X-ray background. One application is
the search for axions and chameleons, where a main background stems from
cosmic rays passing through the detector. The CAST experiment at CERN
uses the helioscope technique by pointing a strong magnet towards the Sun and
searching with various detectors for new particles, which convert into X-ray photons inside the
magnetic field. CAST is currently setting stringent
limits on both axion~\cite{anastassopoulos2017} and Solar chameleon
searches~\cite{anastassopoulos2015}. To further improve the sensitivity, we have shown
that exploiting topological features of a signal is a very
powerful technique and can be performed with a GridPix detector~\cite{krieger}. 
We have, therefore, studied the features of X-ray events of various energies
with the help of an X-ray gun setup at the CAST Detector Lab at CERN~\cite{vafeiadis2012}.

\subsection{X-ray Generation}
The CAST Detector Lab provides a dedicated setup where X-ray photons of various energies can be generated. For this, an electron beam is directed on a
target creating an X-ray spectrum containing the well known characteristic X-ray fluorescence lines of the target material on top of a broad Bremsstrahlung continuum. A dedicated filter is then used to isolate the selected characteristic lines or suppress unwanted parts of the spectrum as good as possible. 
By adjusting the electron beam energy, the target material and the
filter material, quite clean spectra can be created. The settings used are given in
Table~\ref{table_Xray_setting}. The photons are then
guided through a vacuum pipe to the detector. The maximum rate of X-ray photons entering the detector is limited by the filters and windows that were used and by the maximum beam current tolerable for the passive cooled target. This limitation in the X-ray photon rate only affects energies below \SI{2}{\keV} while the detector's readout time of \SI{25}{\milli\second} affects all energies as the readout rate is limited to \SI{40}{\hertz} resulting in a fairly low duty cycle depending on the chosen acquisition times. Therefore, collecting about \num{10000} photons took more than one hour for each energy setting. 
\begin{table}
\begin{center}
\begin{tabular}{c|c|c|c|c}
setup&beam energy&target material&filter&fluorescence line(s)\\
\hline
A&\SI{15}{\keV}&copper&nickel& Cu K$_{\alpha}$ (\SI{8.048}{\keV})\\
\hline
B&\SI{12}{\keV}& manganese & chromium & Mn K$_{\alpha}$ (\SI{5.899}{\keV})\\
\hline
C&\SI{9}{\keV} & titanium & titanium & Ti K$_{\alpha}$ (\SI{4.511}{\keV})  \\
&&&& Ti K$_{\beta}$ (\SI{4.932}{\keV}) \\
\hline
D&\SI{6}{\keV}& silver& silver & Ag L$_{\alpha}$ (\SI{2.984}{\keV})\\
&&&& Ag L$_{\beta}$ (\SI{3.151}{\keV}) \\
\hline
E&\SI{4}{\keV}& aluminum & aluminum & Al K$_{\alpha}$ (\SI{1.487}{\keV}) \\
\hline
F&\SI{2}{\keV}& copper & EPIC & Cu L$_{\alpha}$ (\SI{0.930}{\keV}) \\
&&&& Cu L$_{\beta}$ (\SI{0.950}{\keV}) \\
\hline
G&\SI{0.9}{\keV}& copper & EPIC & O K$_{\alpha}$ (\SI{0.525}{\keV})\\
\hline
H&\SI{0.6}{\keV}& carbon & EPIC & C K$_{\alpha}$ (\SI{0.277}{\keV})\\
\end{tabular}
\caption{Beam energies, target and filter materials chosen to produce photons
  of the listed fluorescence lines by an X-ray generator. A letter is assigned to each setup for reference throughout this document. For some settings more than one fluorescence line is listed, in these cases there was no adequate filter material available to suppress the unwanted line, e.g. the K$_{\beta}$ lines of several target elements. The filter material EPIC is a composite filter composed of a \SI{330}{\nano\metre} thick polypropylene carrier sandwiched by two \SI{90}{\nano\metre} aluminum layers with a \SI{35}{\nano\metre} tin layer on top of one side, also known as \textit{Thick filter} and developed as UV filter for the European Photon Imaging Cameras utilized in the XMM-Newton satellite~\cite{villa1997}. The energies of the lines were taken from the X-ray data booklet~\cite{xdb}.}
\label{table_Xray_setting}
\end{center}
\end{table}

\subsection{Detector}

The detector is described in detail in reference~\cite{nim2017}. A short
summary of the important features is given here. The
readout is a based on a GridPix, which consists of a Timepix ASIC~\cite{llopart2007} on
top of which a Micromegas gas amplification stage (InGrid) is built by
photolithographic post-processing techniques~\cite{chefdeville2006,vandergraaf2007}. The good alignment of each mesh
hole with one readout pixel of the ASIC and low charge threshold of the pixel are features of the
setup: If a primary electron enters in one mesh hole, the electron avalanche
of the gas amplification is collected by a single readout pixel with a typical
threshold of about \num{700} electrons. Thus, each primary electron can be detected with
a very high efficiency~\cite{ottnad2014}, if primary electrons do not end up on the grid and
diffusion spreads the charge cloud sufficiently, so that multiple electrons do
not enter the same mesh hole. 

\begin{figure}
\begin{center}
\subfloat[]{\label{fig:x-ray-efficiency-highE}\includegraphics[width=.8\textwidth]{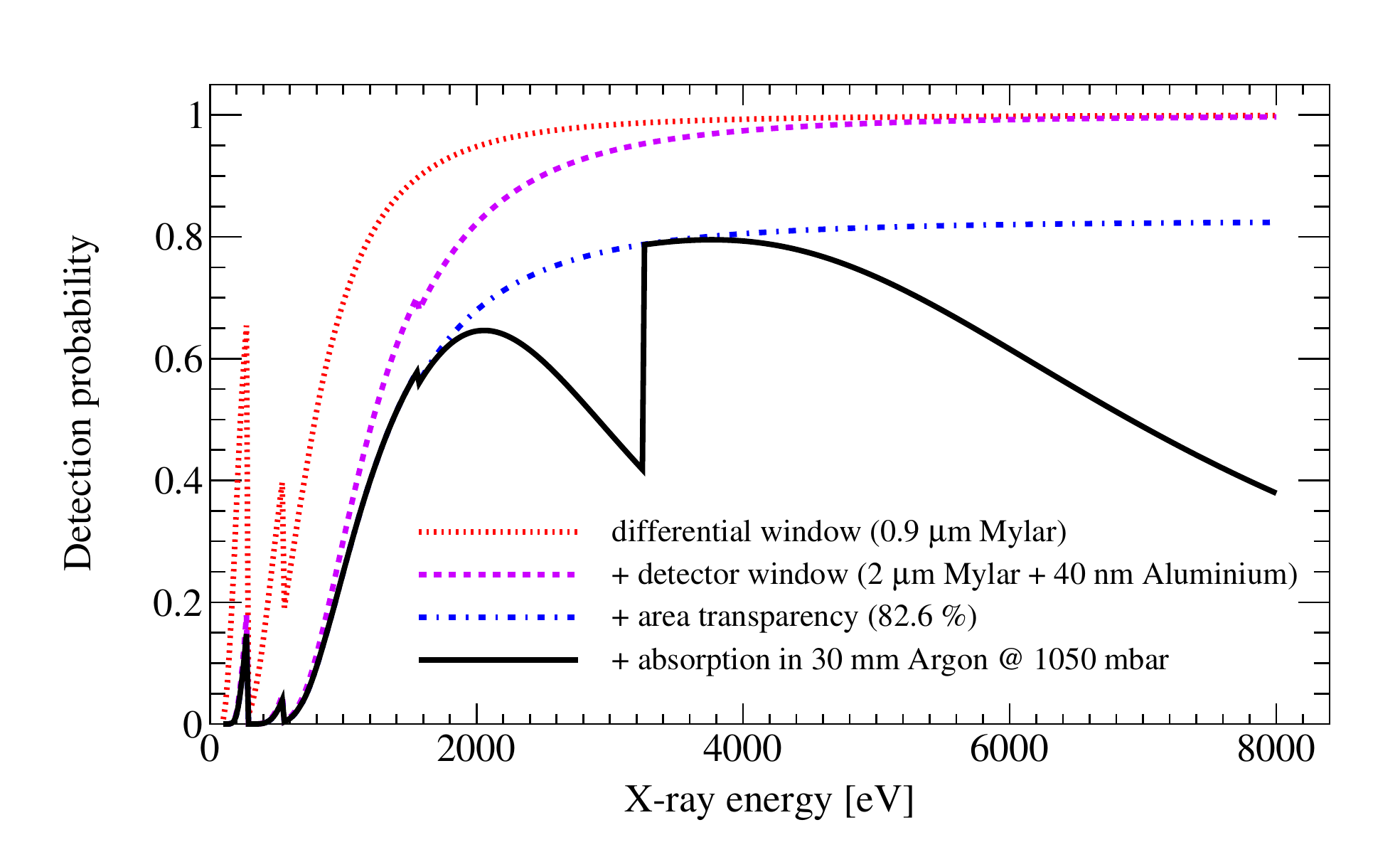}}

\subfloat[]{\label{fig:x-ray-efficiency-lowE}\includegraphics[width=.8\textwidth]{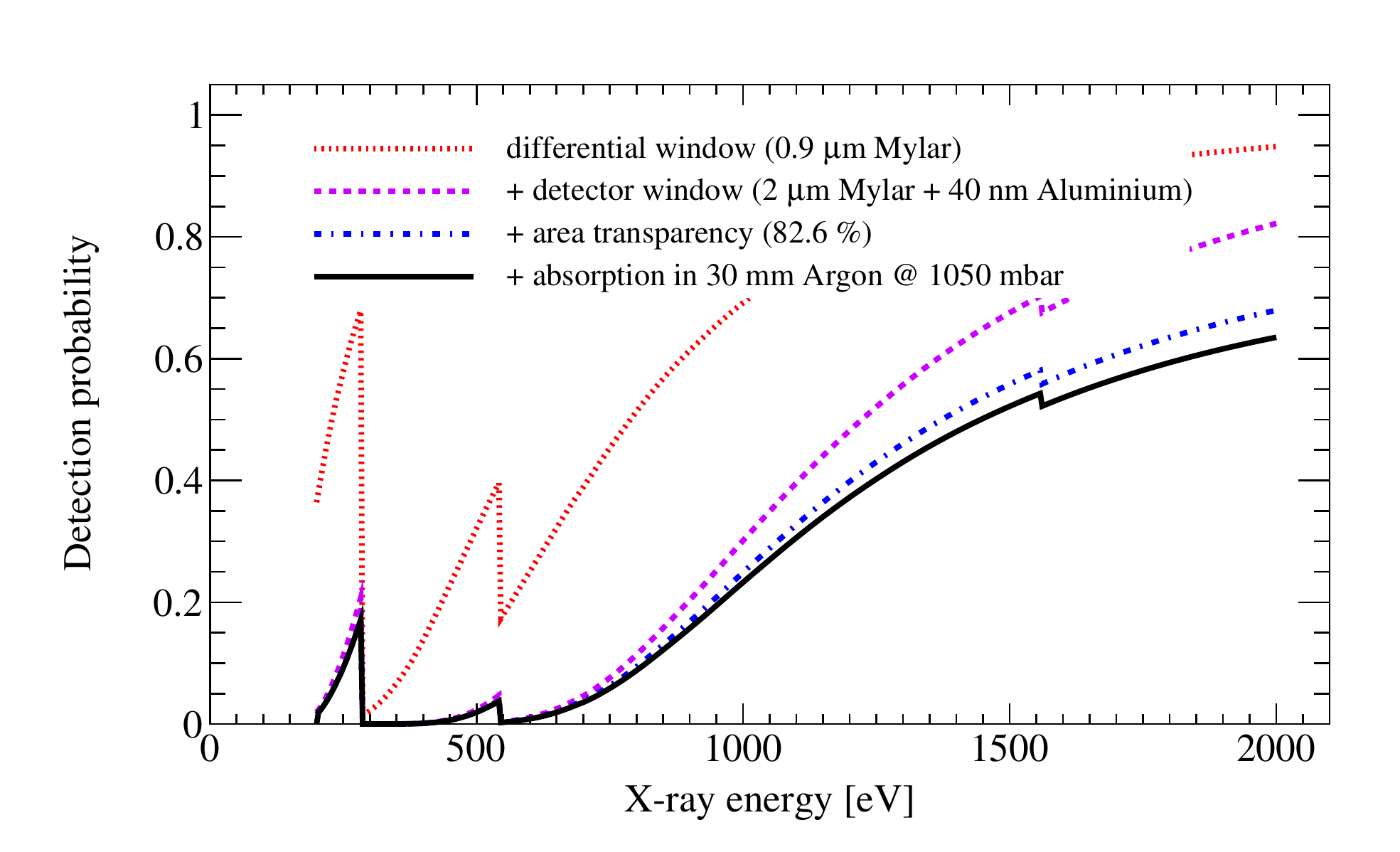}}
\end{center}
\caption{The graph in \protect\subref{fig:x-ray-efficiency-highE} shows the transmission probability as a function of
  the X-ray energy through the differential window, the transmission through 
  the differential and detector window, transmission through
  the differential and detector window including the area transparency of the
  window strong back and the final detection probability including the
  previous transmissions and the absorption in argon. \protect\subref{fig:x-ray-efficiency-lowE} shows a detailed
  view of the low energy range. Transmission and absorption data have been
  produced by a generator using the semi-empirical approach described in
 ~\cite{henke1993}. Graphs and caption taken from~\cite{nim2017}.} 
\label{fig:x-ray-efficiency}
\end{figure}

The drift volume has a maximum drift distance of \SI{3}{\centi\metre} and is filled with an
argon based mixture containing \SI{2.3}{\percent} isobutane as quencher. A drift field of \SI[per-mode=symbol]{500}{\volt\per\centi\metre} is applied. The cathode is made
of a solid \SI{3}{\milli\metre} copper plate, where $\num{5}\times\num{5}$ windows of $\num{3}\times\SI{3}{\milli\metre\squared}$
each have been cut out. The optical transparency of this strong-back is \SI{82.6}{\percent}. A
\SI{2}{\micro\metre} thick Mylar foil with a \SI{40}{\nano\metre} layer of aluminum was
glued on the copper to achieve gas tightness between the drift volume and
the vacuum of the X-ray generator. A differential pumping is, however, still
necessary and requires an additional window consisting of a \SI{0.9}{\micro\metre} thick
Mylar foil separating the good vacuum of the X-ray generator $\left(p\approx\SI{2e-6}{\milli\bar}\right)$ from the bad vacuum $\left(p<\SI{5e-4}{\milli\bar}\right)$ close to the detector. The transmission
probability through the two windows and the detection efficiency as a function
of the X-ray energy are shown in Fig.~\ref{fig:x-ray-efficiency}.

\subsection{Analysis} \label{sub_sec_analysis}
\begin{figure}
\centering
\subfloat[]{\label{fig_xray_277ev}\includegraphics[width=.48\textwidth]{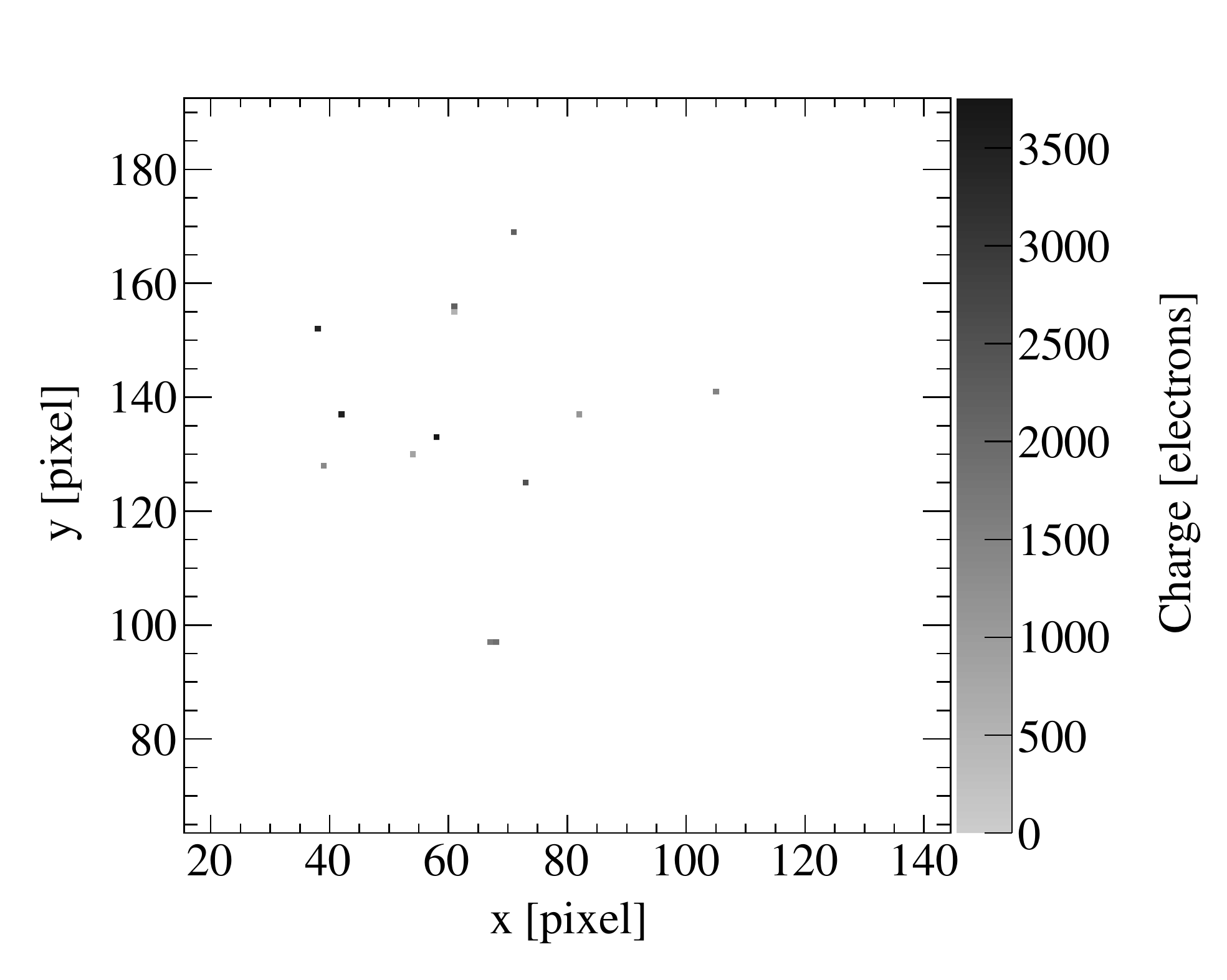}}
\subfloat[]{\label{fig_xray_8kev}\includegraphics[width=.48\textwidth]{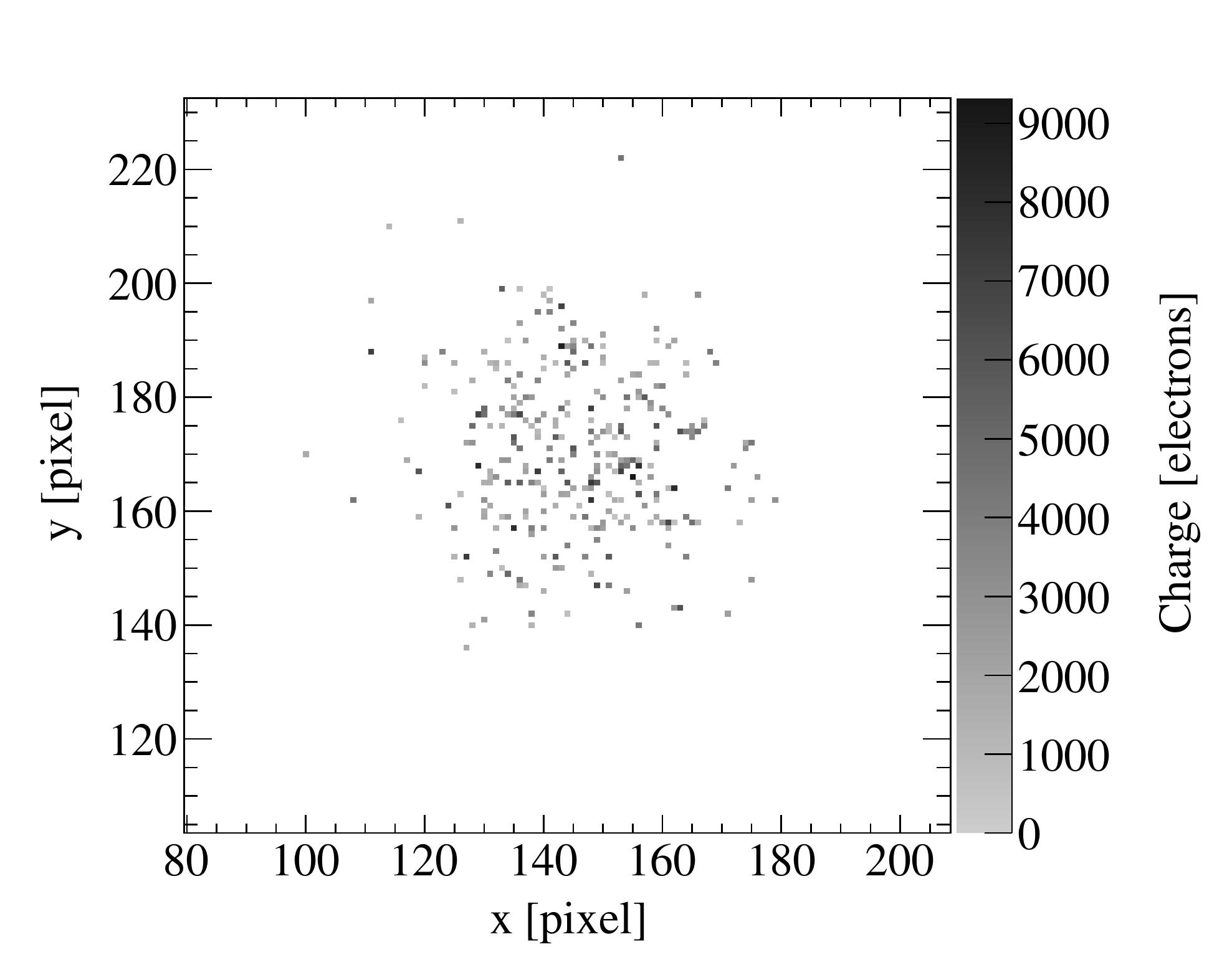}}
\caption{The two images show clusters of a \SI{277}{\eV} X-ray photon \protect\subref{fig_xray_277ev} and of
  an \SI{8}{\keV} X-ray photon \protect\subref{fig_xray_8kev}. Both pictures show only an enlargement of the event
  and not the full sensitive area.}
\label{fig_xray}
\end{figure}

X-ray events are recorded through an unbiased frame-based acquisition. A logical shutter opens for \SI{600}{\micro\second} and then the whole Timepix ASIC is readout. For each hit pixel, the time-over-threshold (ToT) is recorded in a 14-bit pseudo-random counter and used as measure of the collected charge by applying a charge calibration. From the recorded event samples, clean X-ray events are selected employing a number of offline selection cuts.
A first cut requires at least three activated pixels in the
event to reject empty frames. The remaining events are then analyzed using the MarlinTPC software framework~\cite{abernathy2008}. Here
the X-ray photons are reconstructed by searching for a first seed pixel
starting at the top left corner. Additional electrons are assigned to the X-ray
 by searching in a \num{50} by \num{50} pixels array (each pixel measures $\num{55}\times\SI{55}{\micro\metre\squared}$), around
each pixel already found. For every pixel assigned to the
X-ray the same search will be performed in its vicinity. In this
way all pixels belonging to one X-ray photon are identified and X-ray photons far apart from each other are reconstructed
separately. Fig.~\ref{fig_xray} shows the charge clouds of two X-ray photons:
one with an energy of \SI{277}{\eV} and one with \SI{8}{\keV}.

Several properties of each charge cloud are determined. The position of the
X-ray conversion is determined by calculating the mean of all pixel
positions in $x$ and $y$ direction. The mean is required to be less than \SI{4.5}{\milli\metre} from the center of the active area. This ensures a minimal distance of \SI{2.5}{\milli\metre} to the edges, so that even with maximal charge cloud size of
$\sigma(\SI{3}{\centi\metre})=D_T\cdot \sqrt{z}= \SI[per-mode=fraction]{470}{\micro\metre\per\sqrt{\centi\metre}}\cdot\sqrt{\SI{3}{\centi\metre}}\approx \SI{815}{\micro\metre}$ almost no electrons are lost outside of
the sensitive area. Here $D_T$ is the transverse diffusion coefficient for the argon based gas mixture used.

The energy was determined by two different methods: In the first
approach the number of activated pixels gives a very good estimate of the
number of primary electrons, which multiplied by the average ionization energy $W_I$
yields the energy of the X-ray photon. In the second approach, the total charge
$Q$ can be determined by summing over the charge collected by all pixels
assigned to the X-ray photon. The total charge $Q$ is then used as a measure of the energy of the X-ray
photon times the gas gain $G$ and divided by the ionization energy $W_I$.

The spatial width of the charge clouds is slightly asymmetric. This is mostly
because of statistical fluctuations in the diffusion process and can be
measured by the eccentricity of the event. For higher X-ray energies, also
the length of the track of the photoelectron ejected from the atom with an energy exceeding
the binding energy contributes to the asymmetry (e.g. a \SI{5}{\keV} electron has a range of about \SI{500}{\micro\metre} in argon/isobutane \SI{97.7}{\percent}/\SI{2.3}{\percent} at \SI{1050}{\milli\bar}). We have, therefore, identified
the longest axis of the charge cloud and perpendicular to this the shortest
axis. The rms of the charge cloud in the direction of the shorter (=
transverse) axis is referred to as $\sigma_\text{trans}$ and indicates the
diffusion, while the rms value of the pixel position projected on the long
event axis gives $\sigma_\text{long}$.

Finally, topological parameters such as longest axis and event shape
variables such as eccentricity and higher central moments (e.g. kurtosis)
are calculated within this event specific coordinate system. These parameters
are correlated with the diffusion and are therefore temperature dependent. To
avoid this dependence the following three parameters were defined as ratios. These parameters can be used as variables to distinguish X-ray events from charged particle background, e.g. in the analysis of CAST data. 
\begin{enumerate}
\item
Fraction $F_{1\sigma_\text{trans}}$ of pixels within radius of \num{1} $\sigma_\text{trans}$ around center
\item
Eccentricity $\epsilon=\sigma_\text{long}/\sigma_\text{trans}$: a measure for the circularity of an event. By construction, $\epsilon$ is always larger
or equal to \num{1}. 
\item
Length $l$ divided by $\sigma_\text{trans}$: The length is defined as the
distance of the outermost points in the projection onto the long event axis.
\end{enumerate}

Due to insufficient matching of the shutter times to the photon rate provided
by the X-ray generator some of the datasets contain a significant fraction of
double events where the two or more X-ray photons cannot be separated by the
algorithm. To filter out these events, loose cuts on eccentricity $\epsilon$, length $l$, and
transverse rms $\sigma_\text{trans}$ where applied, the cuts for each setup are listed in Table~\ref{table_cuts}. In case of $l$ and $\sigma_\text{trans}$ the cuts were chosen such that only events incompatible with the single photon hypothesis were rejected.

\begin{table}
\begin{center}
\begin{tabular}{c|c}
setup&applied cuts\\
\hline
A&$\epsilon < \num{1.3}$, $\SI{0.1}{\milli\metre}<\sigma_\text{trans}\le\SI{1.0}{\milli\metre}$\\
B&$\epsilon < \num{1.3}$, $\SI{0.1}{\milli\metre}<\sigma_\text{trans}\le\SI{1.0}{\milli\metre}$\\
C&$\epsilon < \num{1.3}$, $\SI{0.1}{\milli\metre}<\sigma_\text{trans}\le\SI{1.0}{\milli\metre}$\\
D&$\epsilon < \num{1.4}$, $\SI{0.1}{\milli\metre}<\sigma_\text{trans}\le\SI{1.0}{\milli\metre}$, $l\le\SI{6}{\milli\metre}$\\
E&$\epsilon < \num{2.0}$, $\SI{0.1}{\milli\metre}<\sigma_\text{trans}\le\SI{1.1}{\milli\metre}$\\
F&$\epsilon < \num{2.0}$, $\SI{0.1}{\milli\metre}<\sigma_\text{trans}\le\SI{1.1}{\milli\metre}$\\
G&$\epsilon < \num{2.0}$, $\SI{0.1}{\milli\metre}<\sigma_\text{trans}\le\SI{1.1}{\milli\metre}$\\
H&$\SI{0.1}{\milli\metre}<\sigma_\text{trans}\le\SI{1.1}{\milli\metre}$, $l\le\SI{6}{\milli\metre}$
\end{tabular}
\end{center}
\caption{Cuts on eccentricity $\epsilon$, transverse rms $\sigma_\text{trans}$ and length $l$, applied to the data recorded with the different X-ray gun setups. Cut values are chosen rather loose in order to only reject events incompatible with the single photon hypothesis. Additionaly, for all setups a minimum number of \num{3} active pixels is required to reject empty events. Also, all accepted events are required to have their center of gravity within a \SI{4.5}{\milli\metre} radius around the chip center to avoid events only partially contained in the active area.}
\label{table_cuts}
\end{table}

\section{Spectra and Energy Calibration}\label{sec_spectra}
\begin{figure}
\centering
\subfloat[]{\label{fig_spectra_pixel_A}\includegraphics[width=.45\textwidth]{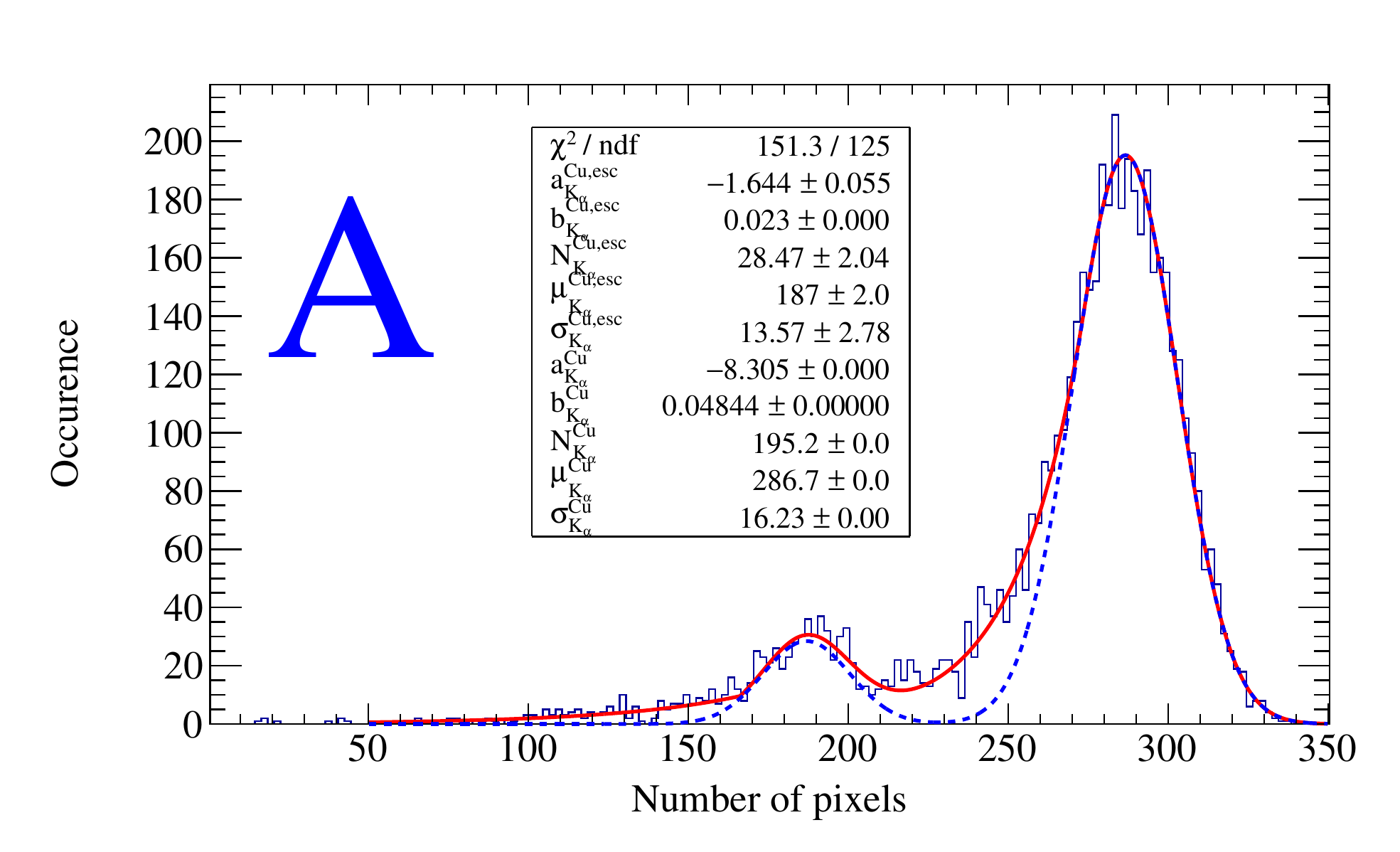}}
\subfloat[]{\label{fig_spectra_charge_A}\includegraphics[width=.45\textwidth]{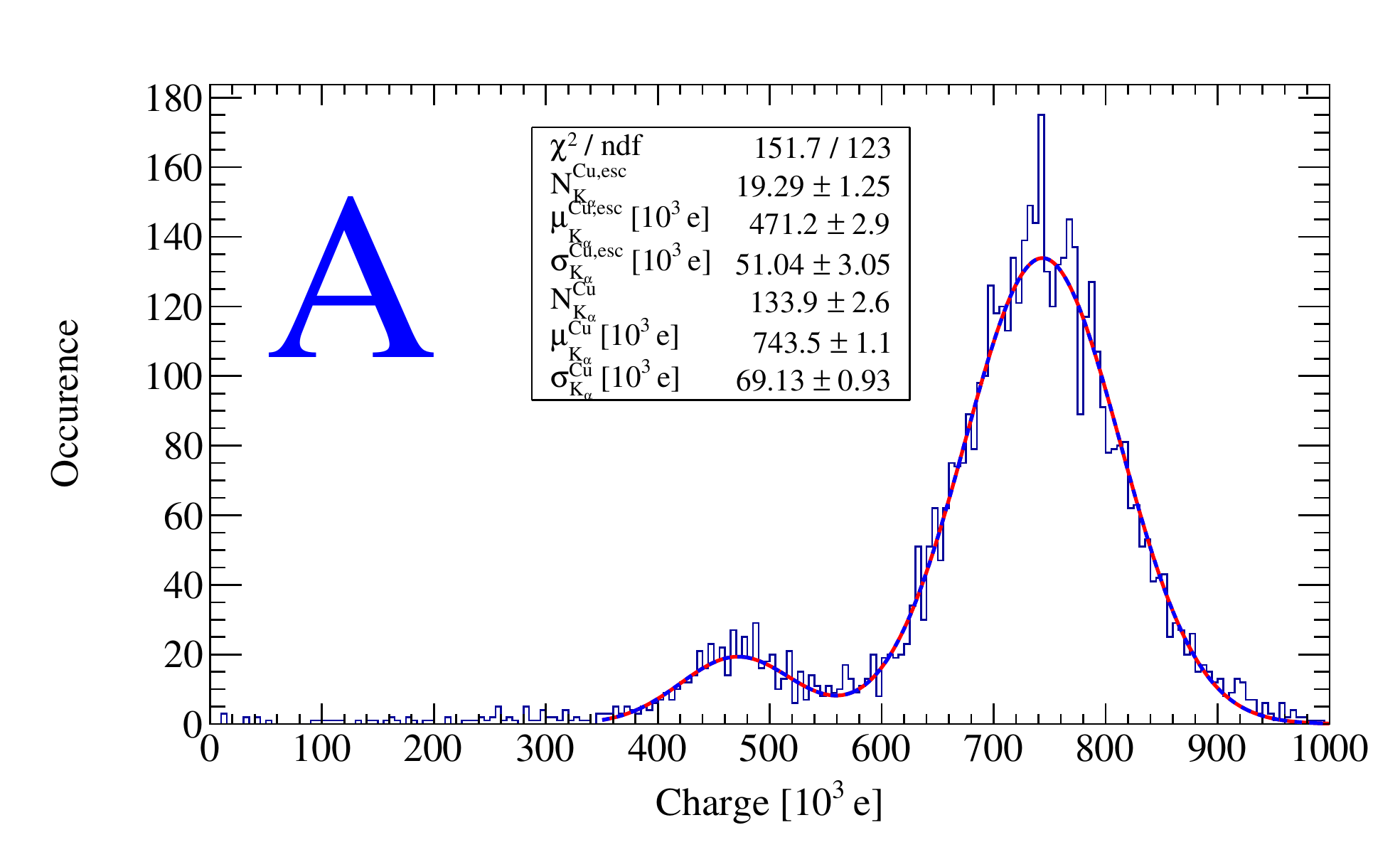}}\\
\vspace*{-.45cm}
\subfloat[]{\label{fig_spectra_pixel_B}\includegraphics[width=.45\textwidth]{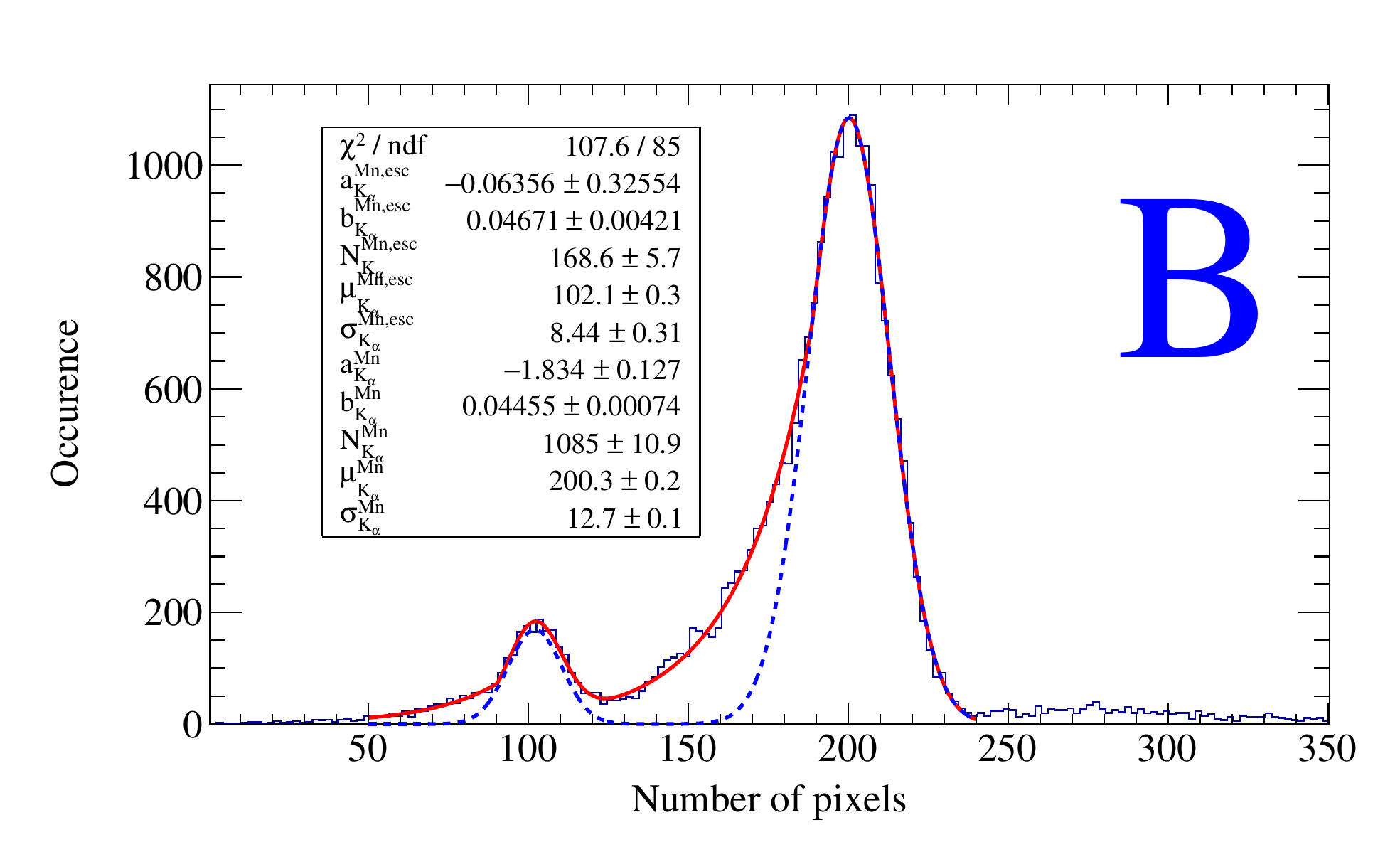}}
\subfloat[]{\label{fig_spectra_charge_B}\includegraphics[width=.45\textwidth]{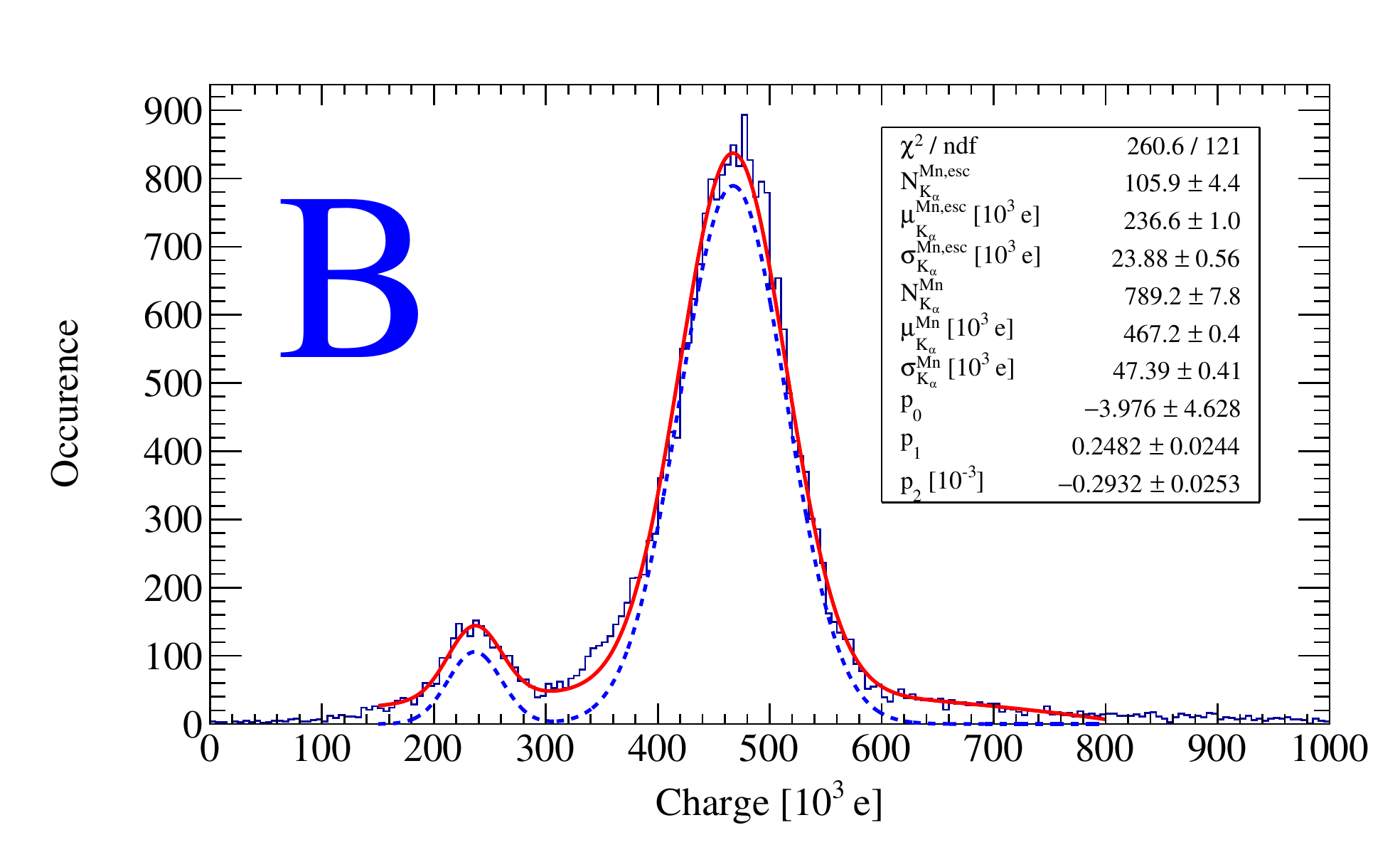}}\\
\vspace*{-.45cm}
\subfloat[]{\label{fig_spectra_pixel_C}\includegraphics[width=.45\textwidth]{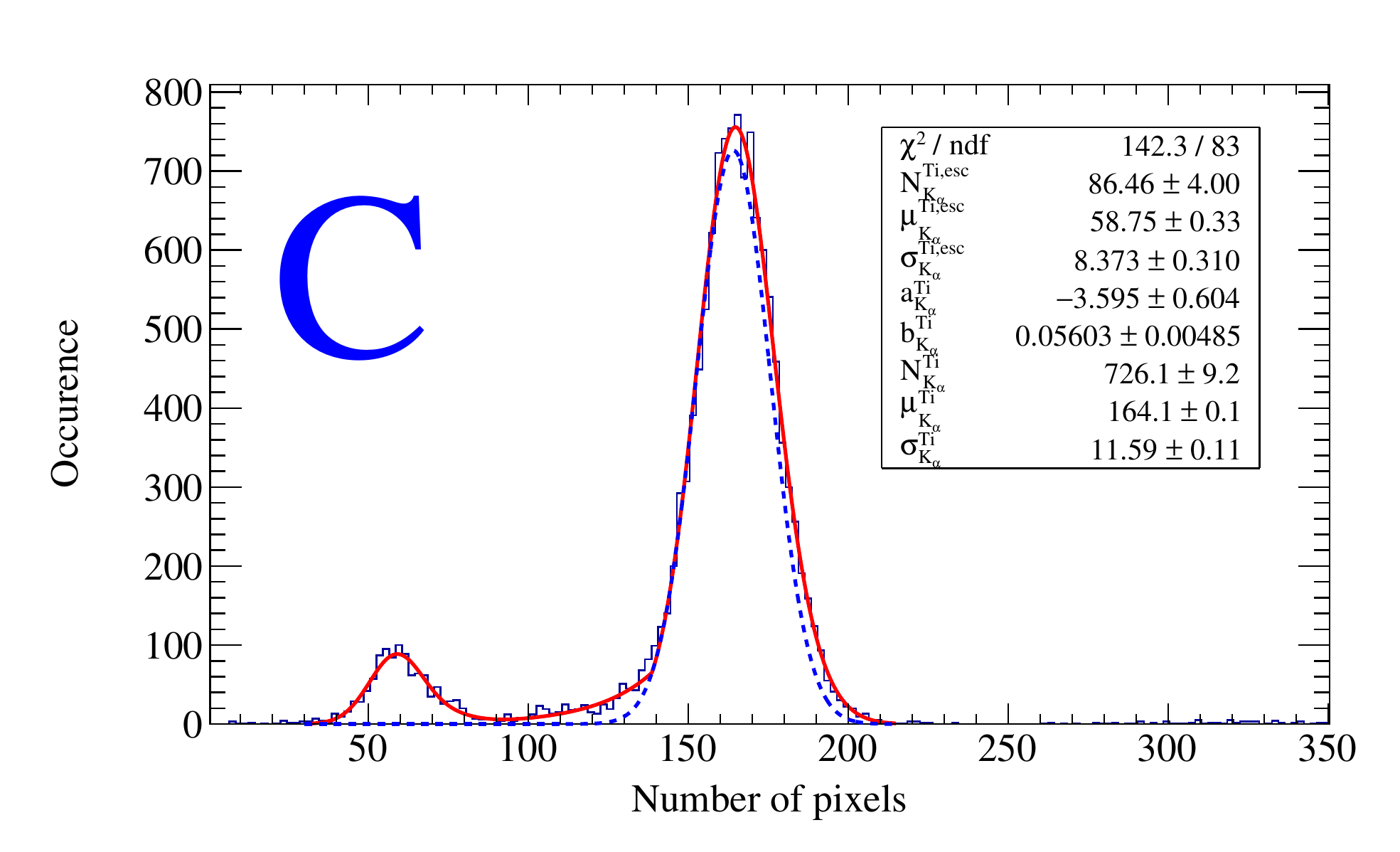}}
\subfloat[]{\label{fig_spectra_charge_C}\includegraphics[width=.45\textwidth]{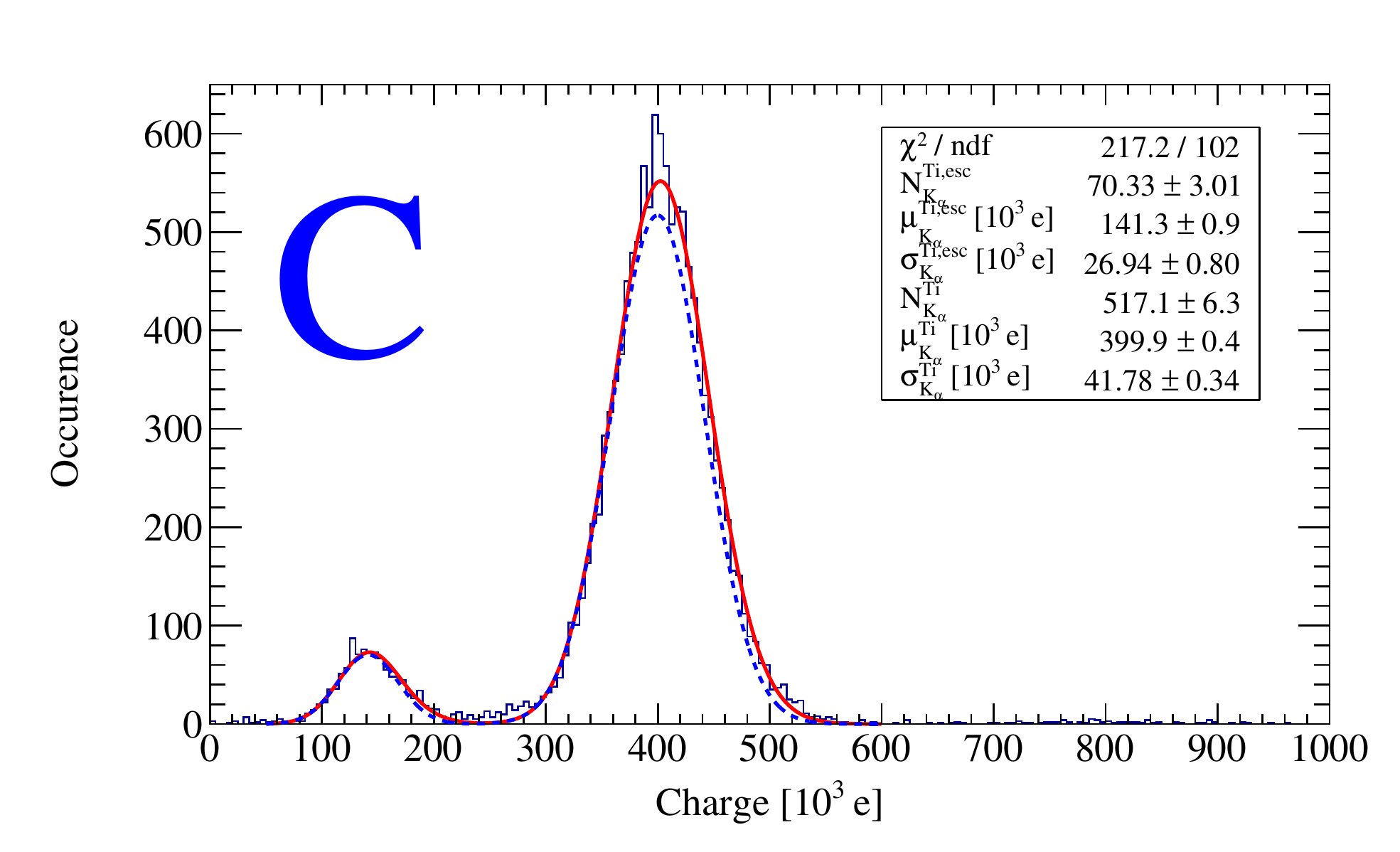}}\\
\vspace*{-.45cm}
\subfloat[]{\label{fig_spectra_pixel_D}\includegraphics[width=.45\textwidth]{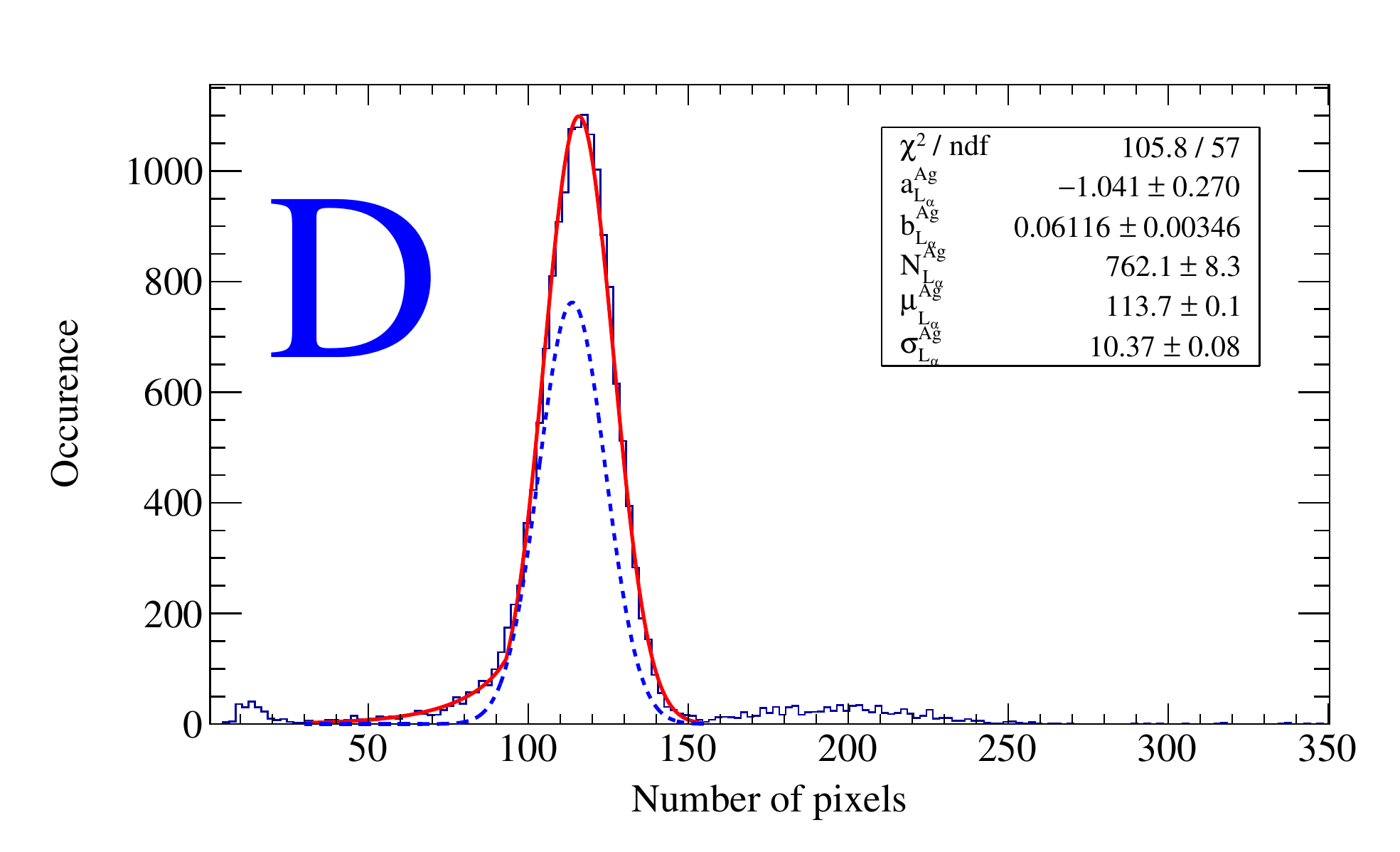}}
\subfloat[]{\label{fig_spectra_charge_D}\includegraphics[width=.45\textwidth]{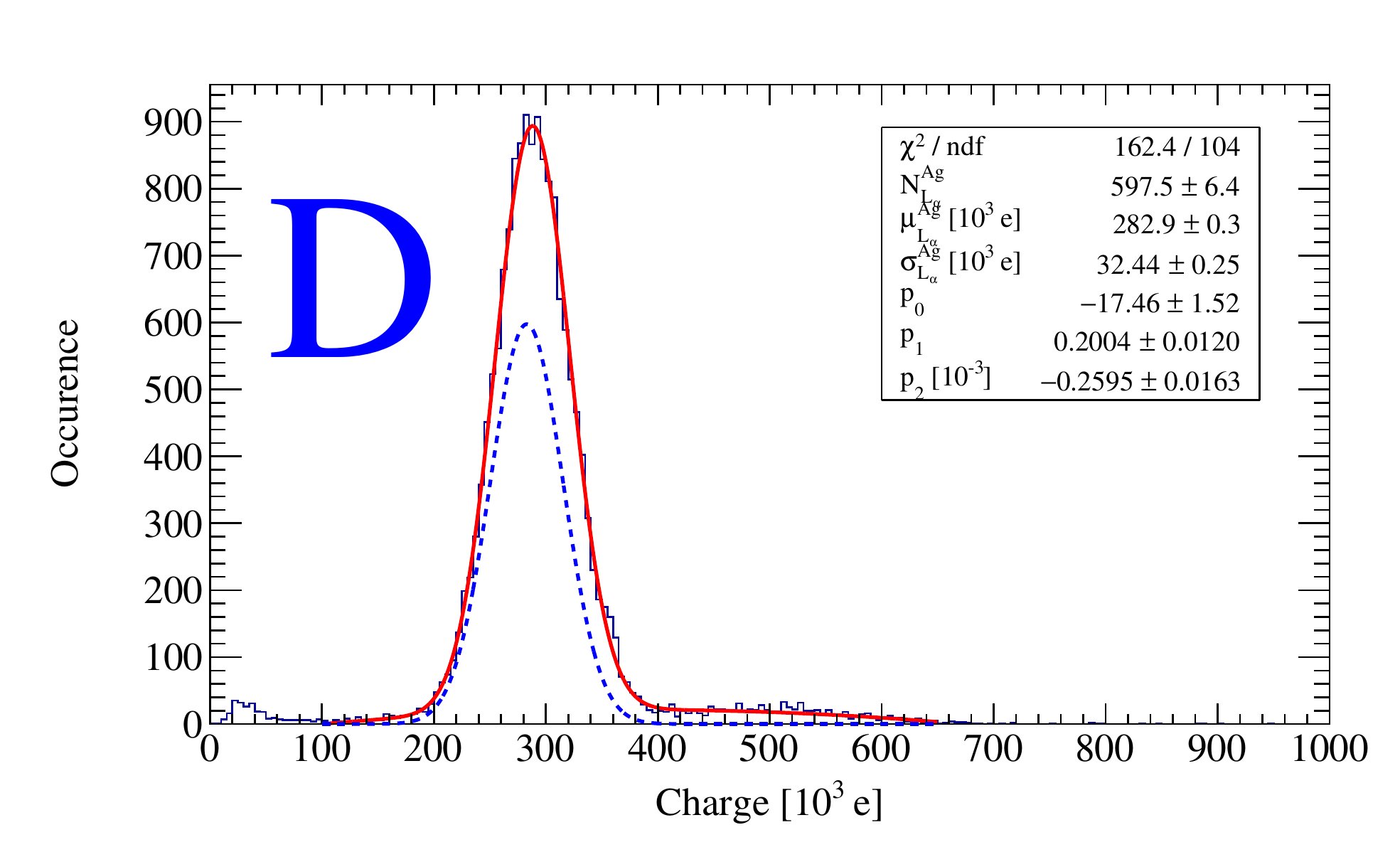}}
\vspace*{-.2cm}
\caption{Energy spectra of various settings based on the number of primary
  electrons (left) and total charge (right). Main peaks shown are the copper K$_\alpha$ line at \SI{8}{\keV}: \protect\subref{fig_spectra_pixel_A} and \protect\subref{fig_spectra_charge_A}; the manganese K$_\alpha$ line at \SI{5.9}{\keV}: \protect\subref{fig_spectra_pixel_B} and \protect\subref{fig_spectra_charge_B}; the titanium K$_\alpha$ line at \SI{4.5}{\keV}: \protect\subref{fig_spectra_pixel_C} and \protect\subref{fig_spectra_charge_C}; and the silver L$_\alpha$ line at \SI{3}{\keV}: \protect\subref{fig_spectra_pixel_D} and \protect\subref{fig_spectra_charge_D}. The functions fitted to the spectra are shown in solid red while the Gaussians describing the main peaks are plotted in addition as blue dashed line.}
\label{fig_spectra1}
\end{figure}

\begin{figure}
\centering
\subfloat[]{\label{fig_spectra_pixel_E}\includegraphics[width=.45\textwidth]{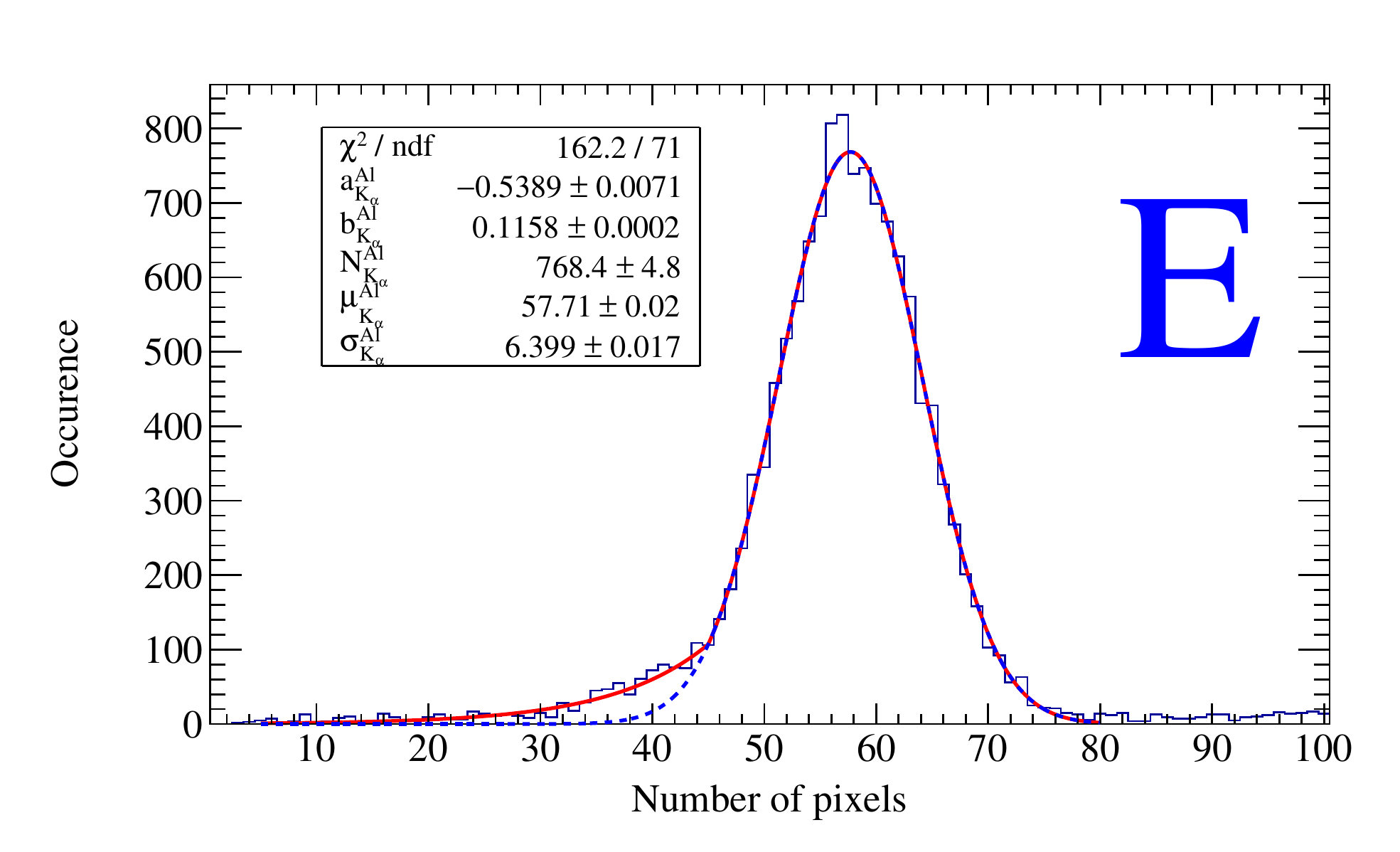}}
\subfloat[]{\label{fig_spectra_charge_E}\includegraphics[width=.45\textwidth]{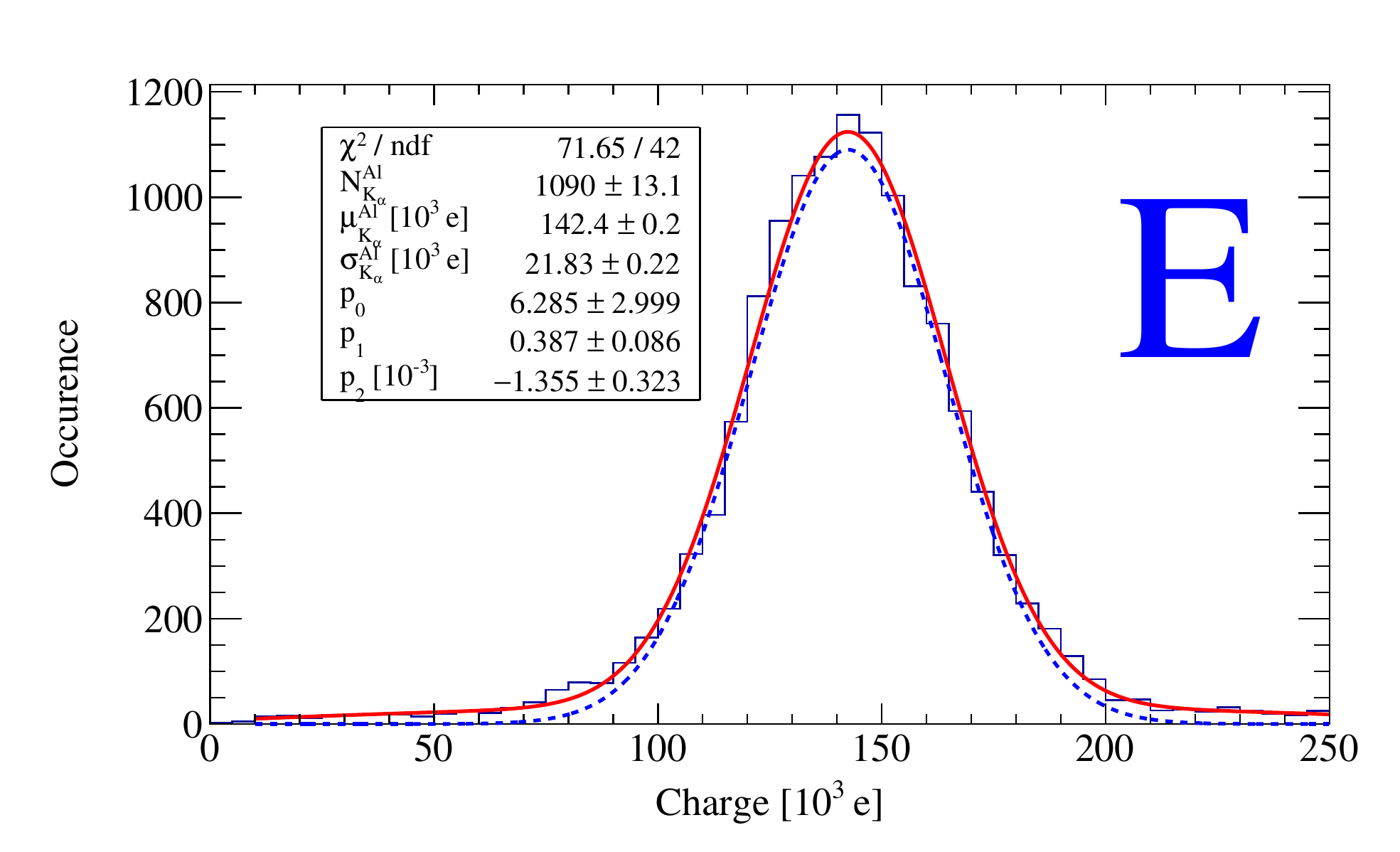}}\\
\vspace*{-.45cm}
\subfloat[]{\label{fig_spectra_pixel_F}\includegraphics[width=.45\textwidth]{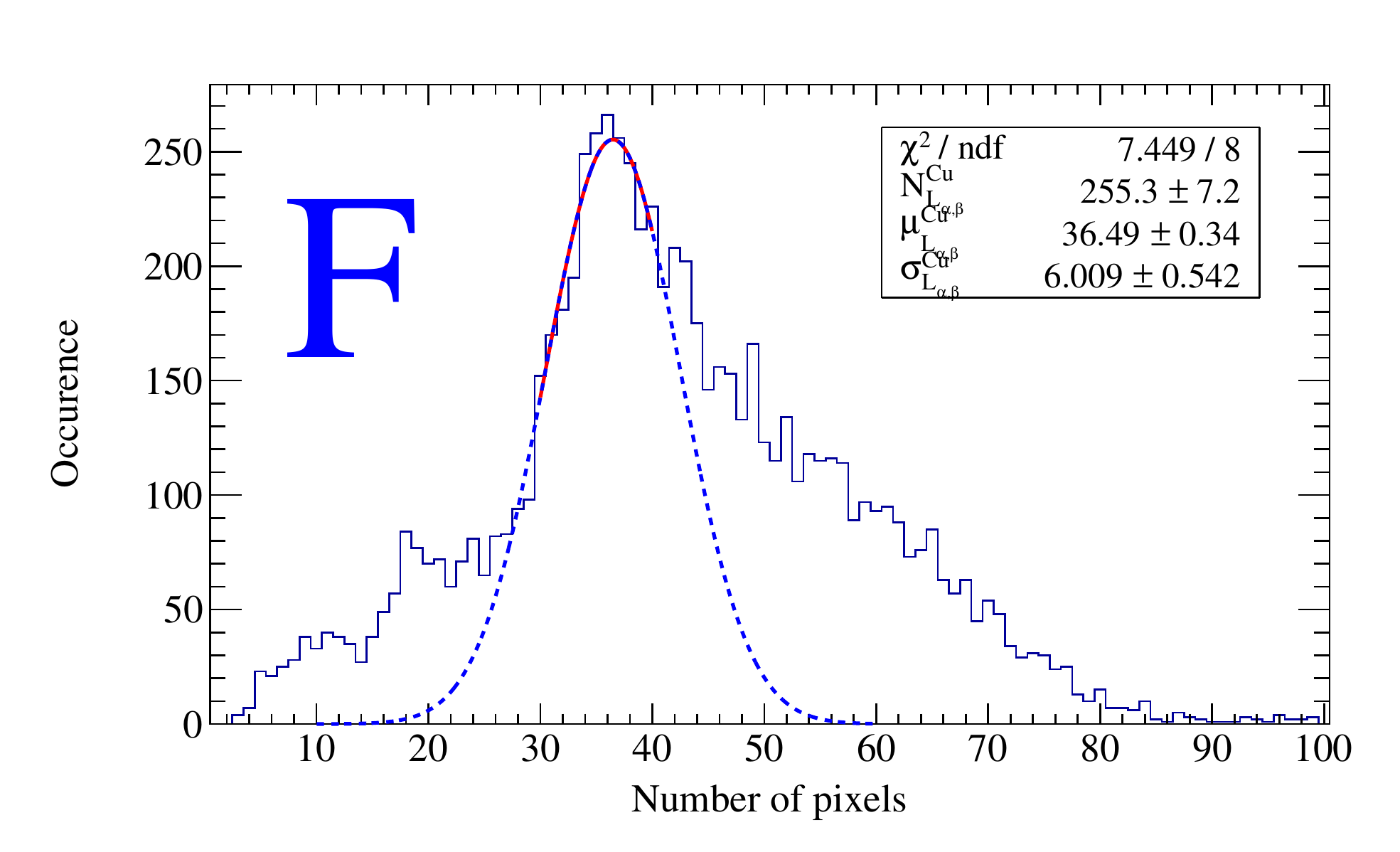}}
\subfloat[]{\label{fig_spectra_charge_F}\includegraphics[width=.45\textwidth]{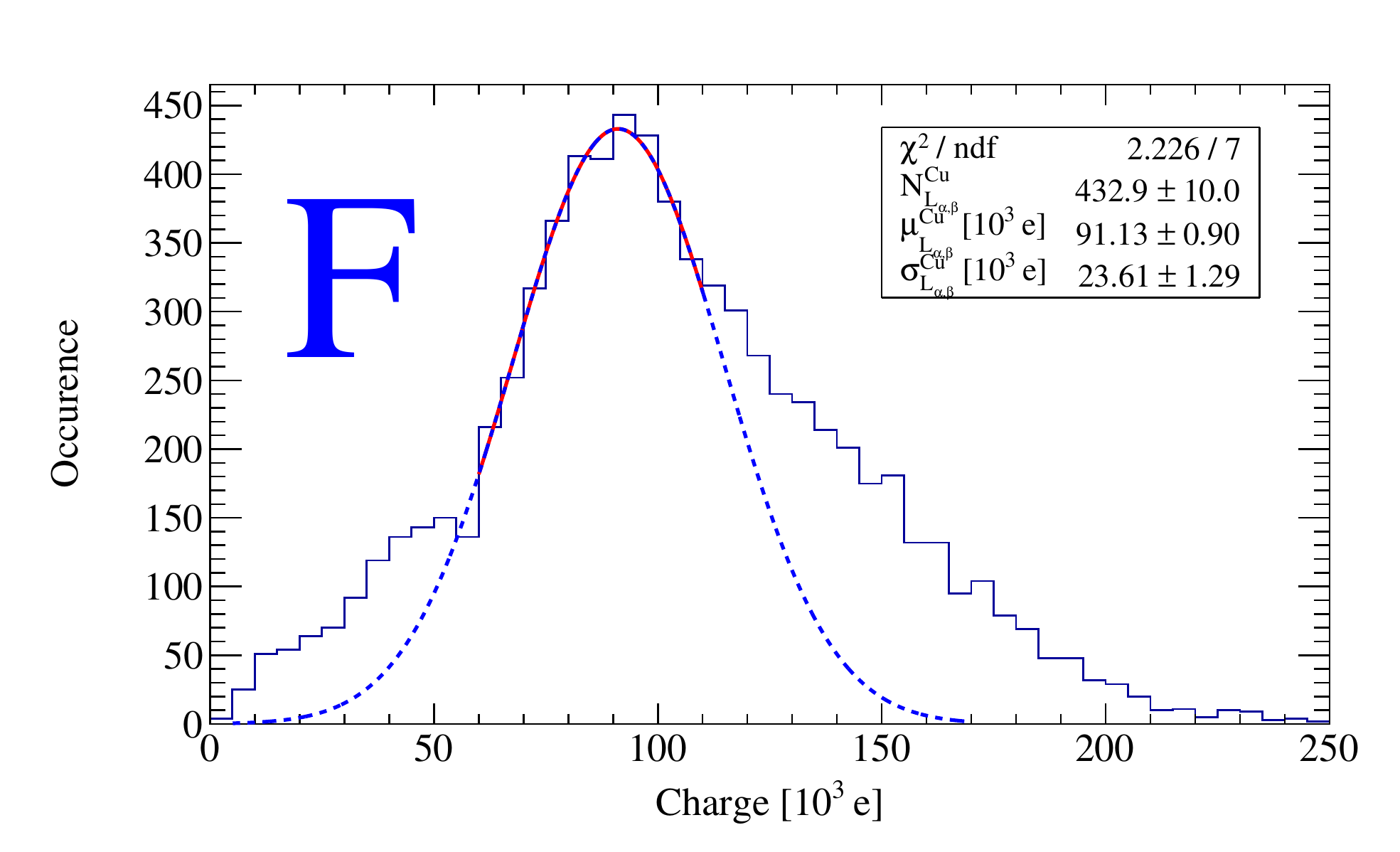}}\\
\vspace*{-.45cm}
\subfloat[]{\label{fig_spectra_pixel_G}\includegraphics[width=.45\textwidth]{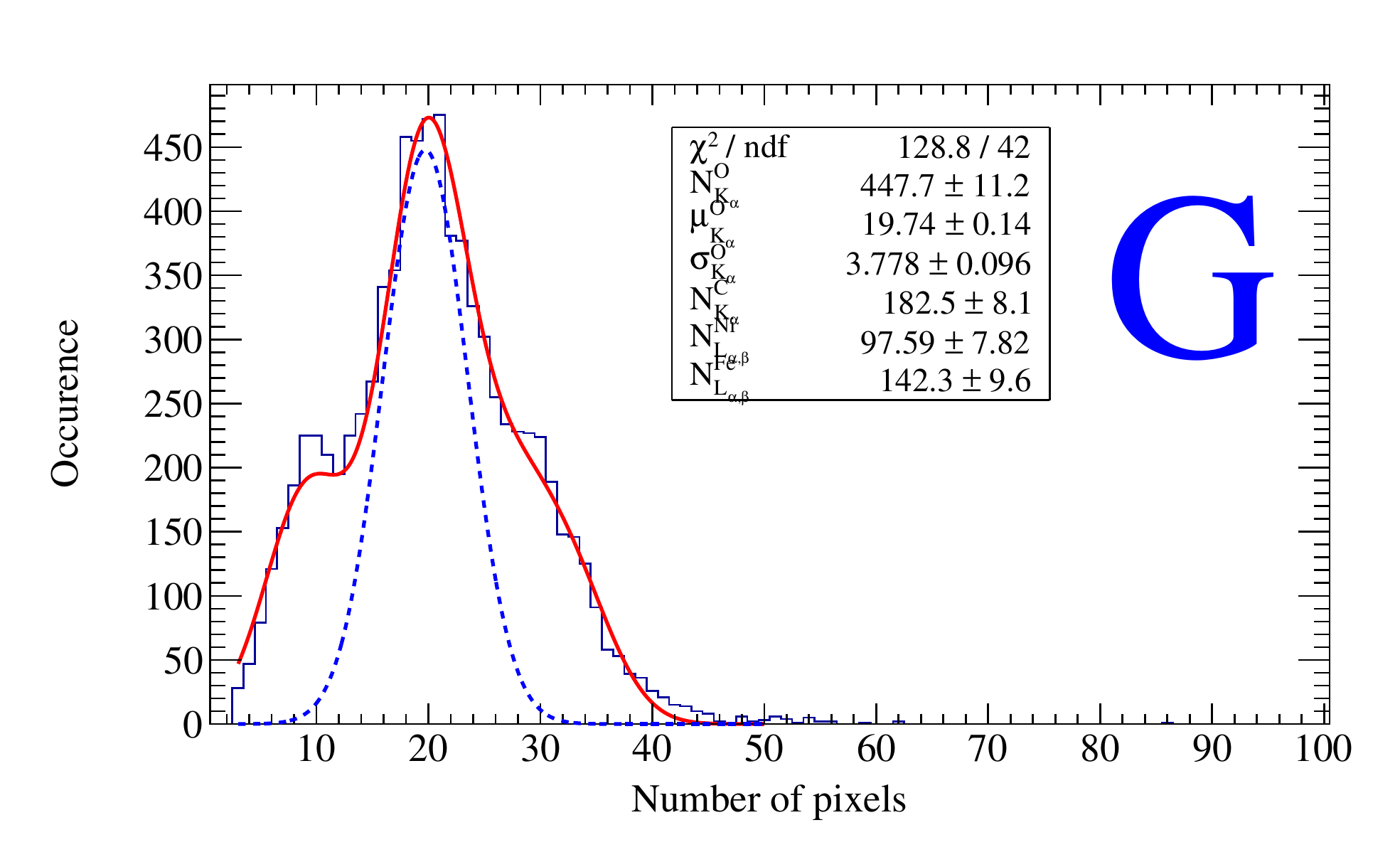}}
\subfloat[]{\label{fig_spectra_charge_G}\includegraphics[width=.45\textwidth]{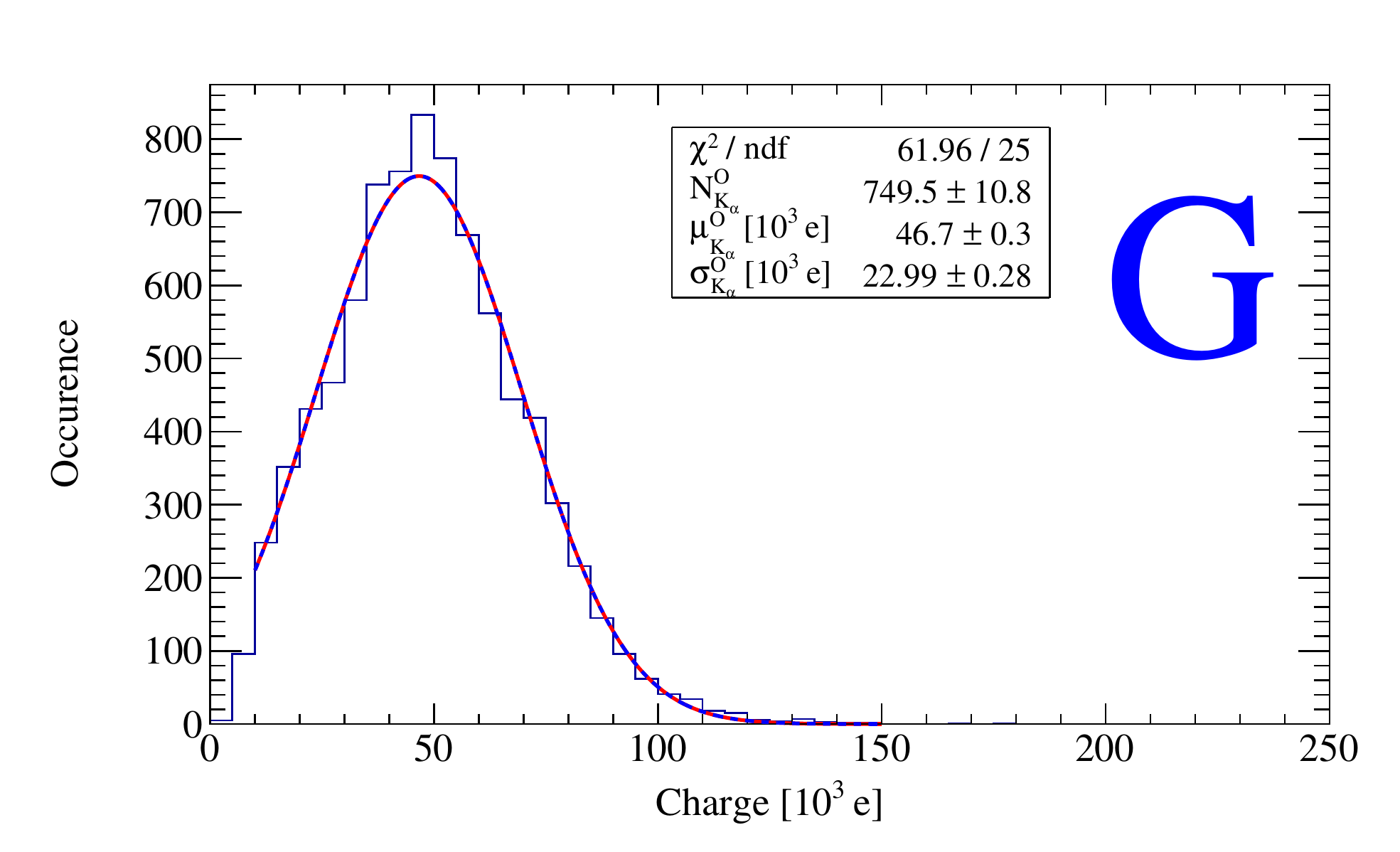}}\\
\vspace*{-.45cm}
\subfloat[]{\label{fig_spectra_pixel_H}\includegraphics[width=.45\textwidth]{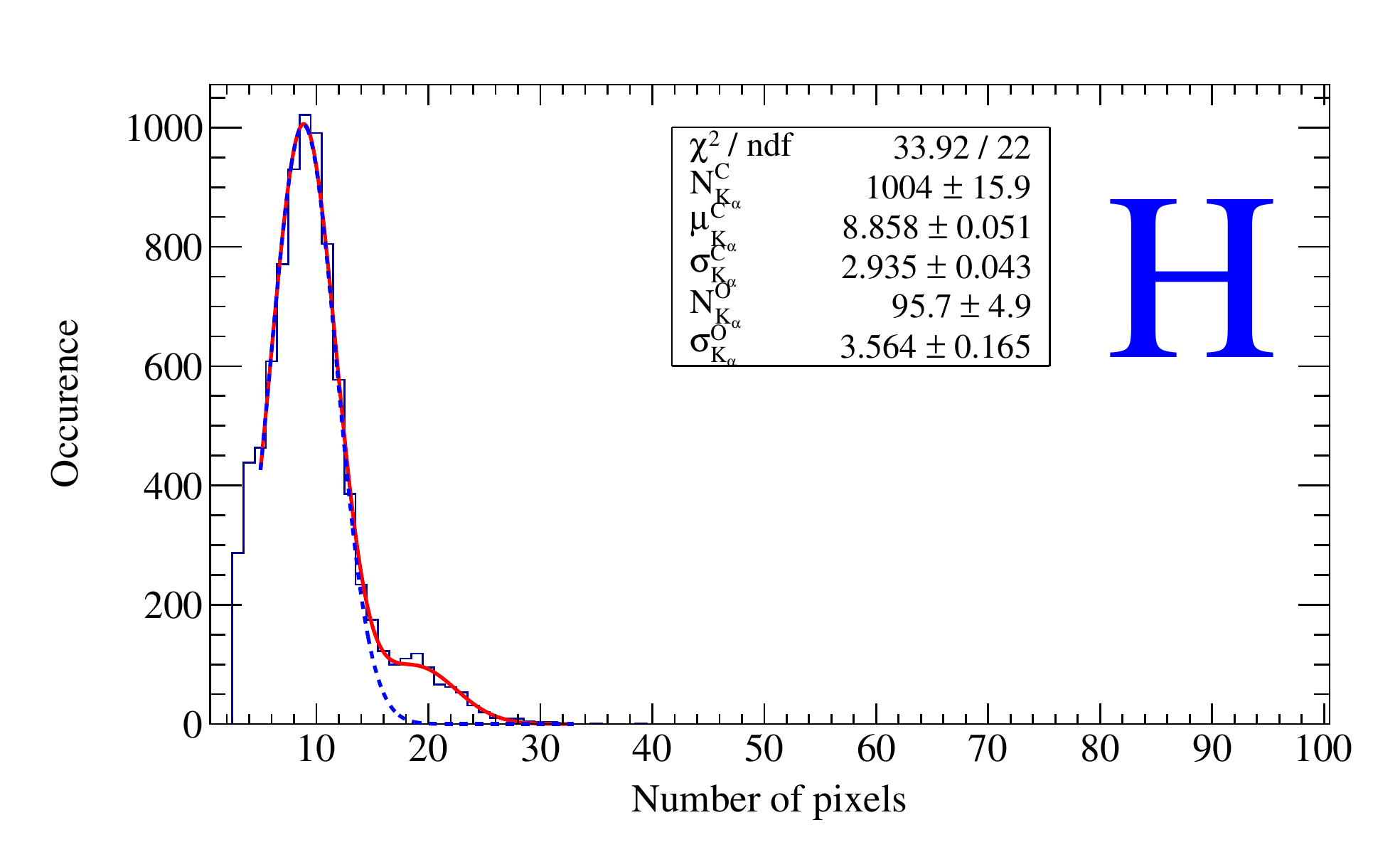}}
\subfloat[]{\label{fig_spectra_charge_H}\includegraphics[width=.45\textwidth]{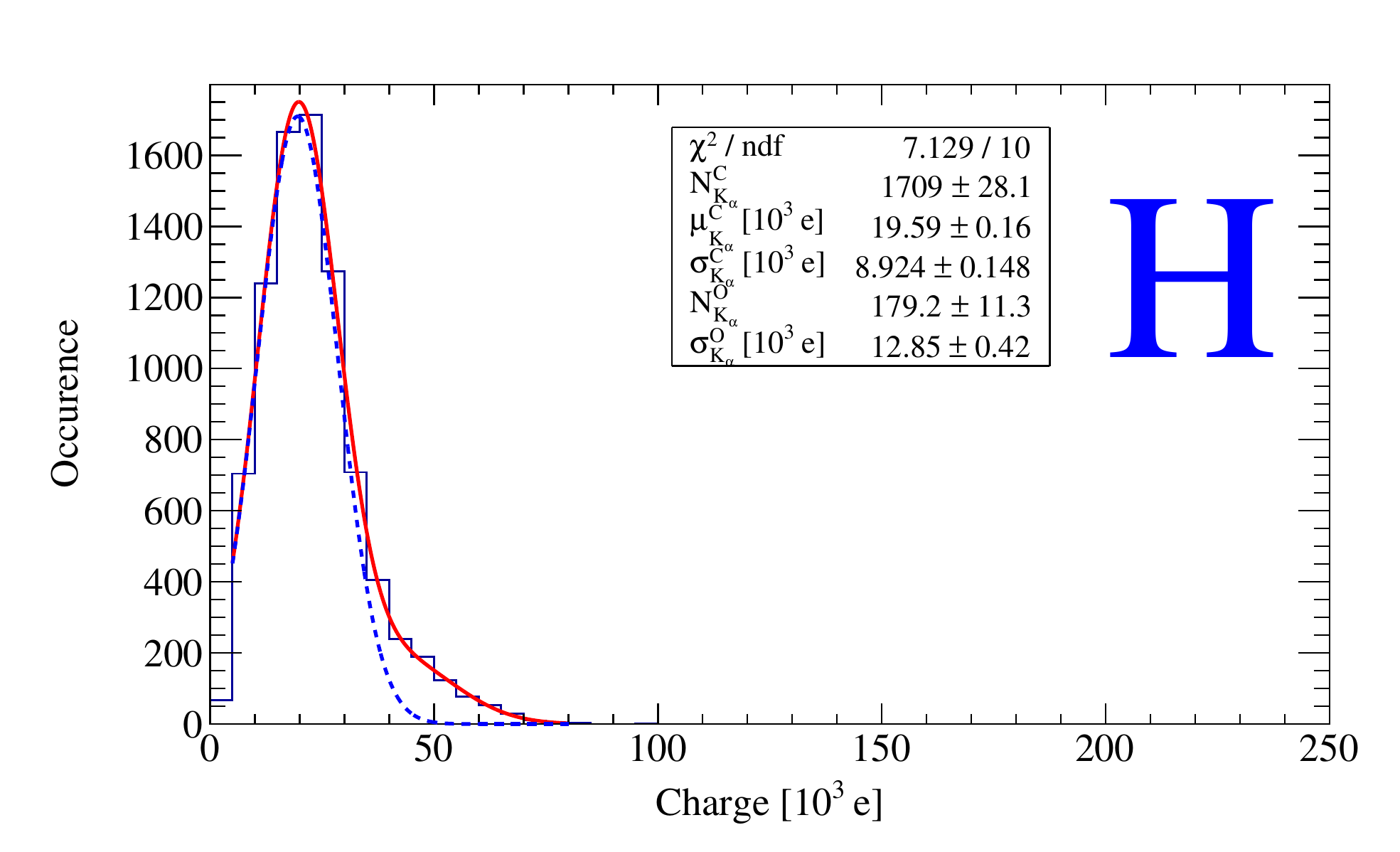}}
\vspace*{-.2cm}
\caption{Energy spectra of various settings based on the number of primary
  electrons (left) and total charge (right). Main peaks shown are the aluminum K$_\alpha$ line at \SI{1.5}{\keV}: \protect\subref{fig_spectra_pixel_E} and \protect\subref{fig_spectra_charge_E}; the copper L$_\alpha$ line at \SI{0.9}{\keV}: \protect\subref{fig_spectra_pixel_F} and \protect\subref{fig_spectra_charge_F}; the oxygen K$_\alpha$ line at \SI{0.5}{\keV}: \protect\subref{fig_spectra_pixel_G} and \protect\subref{fig_spectra_charge_G}; and the carbon K$_\alpha$ line at \SI{277}{\eV}: \protect\subref*{fig_spectra_pixel_H} and \protect\subref{fig_spectra_charge_H}. The functions fitted to the spectra are shown in solid red while the Gaussians describing the main peaks are plotted in addition as blue dashed line.}
\label{fig_spectra2}
\end{figure}

For each electron beam energy, target and filter combination described in
Table~\ref{table_Xray_setting}, the resulting spectrum was created. 
They are shown in Figs.~\ref{fig_spectra1} and~\ref{fig_spectra2}, using the number of primary
electrons (left column) and the total charge (right column) as measure for the detected energy.

Some of the spectra, in addition to the main peak, contain extra peaks due to different processes or causes. For energies above \SI{3}{\keV} the argon escape
line of the main line appears; e.g. in the spectrum of setup C (see Figs.~\ref{fig_spectra_pixel_C} and~\ref{fig_spectra_charge_C}) the argon escape line is visible approximately \SI{3}{\keV} below the main titanium K$_\alpha$ line at \SI{4.5}{\keV}. Especially for the low energies, additional fluorescence lines show up, close to the main line which can be attributed to contaminations of the target material. For example the prominent carbon line (see Figs.~\ref{fig_spectra_pixel_H} and~\ref{fig_spectra_charge_H}) also has a visible shoulder at higher energies corresponding to the oxygen K$_{\alpha}$ line at \SI{525}{\eV}
which is produced by a small oxygen contamination of the target surface. All lines considered in the fits are
listed in table~\ref{table_Xray_lines}.

\begin{table}
\begin{center}
\begin{tabular}{c|c|c}
fluorescence line&additional peaks&fixed parameters\\
\hline
Cu K$_{\alpha}$ (\SI{8.048}{\keV})&Cu K$_{\alpha}$ escape (\SI{5.057}{\keV})&none\\
\hline
Mn K$_{\alpha}$ (\SI{5.899}{\keV})&Mn K$_{\alpha}$ escape (\SI{2.925}{\keV})&none\\
\hline
Ti K$_{\alpha}$ (\SI{4.511}{\keV})&Ti K$_{\beta}$ (\SI{4.932}{\keV})&$\mu^{\text{Ti}}_{\text{K}_\beta}$, $\sigma^{\text{Ti}}_{\text{K}_\beta}$, $N^{\text{Ti}}_{\text{K}_\beta}/N^{\text{Ti}}_{\text{K}_\alpha}$\\
&Ti K$_{\alpha}$ escape (\SI{1.537}{\keV})&\\
&Ti K$_{\beta}$ escape(\SI{1.959}{\keV})&$\mu^{\text{Ti-esc}}_{\text{K}_\beta}$, $\sigma^{\text{Ti-esc}}_{\text{K}_\beta}$\\
\hline
Ag L$_{\alpha}$ (\SI{2.984}{\keV})&Ag L$_{\beta}$ (\SI{3.151}{\keV})&$\mu^{\text{Ag}}_{\text{L}_\beta}$,$\sigma^{\text{Ag}}_{\text{L}_\beta}$,$N^{\text{Ag}}_{\text{L}_\beta}/N^{\text{Ag}}_{\text{L}_\alpha}$\\
\hline
Al K$_{\alpha}$ (\SI{1.487}{\keV})&none&none\\
\hline
Cu L$_{\alpha}$ (\SI{0.930}{\keV})&see note in caption&see note in caption\\
\hline
O K$_{\alpha}$ (\SI{0.525}{\keV})&C K$_{\alpha}$ (\SI{0.277}{\keV})&$\mu^{\text{C}}_{\text{K}_\alpha}$,$\sigma^{\text{C}}_{\text{K}_\alpha}$\\
&Fe L$_{\alpha\mathrm{,}\beta}$ (\SI{0.71}{\keV})&$\mu^{\text{Fe}}_{\text{K}_{\alpha\mathrm{,}\beta}}$,$\sigma^{\text{Fe}}_{\text{K}_{\alpha\mathrm{,}\beta}}$\\
&Ni L$_{\alpha\mathrm{,}\beta}$ (\SI{0.86}{\keV})&$\mu^{\text{Ni}}_{\text{K}_{\alpha\mathrm{,}\beta}}$,$\sigma^{\text{Ni}}_{\text{K}_{\alpha\mathrm{,}\beta}}$\\
\hline
C K$_{\alpha}$ (\SI{0.277}{\keV})&O K$_{\alpha}$ (\SI{0.525}{\keV})&$\mu^{\text{O}}_{\text{K}_\alpha}$,$\sigma^{\text{O}}_{\text{K}_\alpha}$
\end{tabular}
\caption{X-ray lines visible in the different spectra. For some of the spectra additional lines have to be taken into account in the fitted functions which stem from argon escape lines, close-by $\beta$ lines and/or additional elements present in the target material (possible contaminations). To simplify the fits as many parameters as possible have been fixed for the additional lines: The mean $\mu$ of the fitted Gaussian is usually fixed relative to the position of the main peak while the width $\sigma$ is assumed to be the same as for the close-by main peak. For $\beta$ lines also the relative intensity is used to fully fix the additional peak through the main peak. In case of the copper L$_{\alpha}$ line there are definitely contributions by other X-ray lines visible in the spectrum but no neighboring peaks could be clearly identified, therefore in this case the fit range was narrowed to the main peak. In case of the oxygen K$_{\alpha}$ line many contaminants show up, possibly present in the form of stainless steel screws used to mount the target. Additional peaks were identified using the tabulated X-ray fluorescence energies in~\cite{xdb} from which also the information used to fix some of the fit parameters were taken.}
\label{table_Xray_lines}
\end{center}
\end{table}

The main peak is always described by a Gaussian function with three free
parameters: amplitude of the Gaussian $N$, position of mean $\mu$ and width
$\sigma$. Some parameters of the side peaks are linked to
values of the main peak. For example, the same width is used, if the side peak
is sufficiently close, or the position is fixed, if the relative or
fractional energy difference is known. The list of the fixed parameters is
also given in Table~\ref{table_Xray_lines}, the full fit functions used are listed in Tables~\ref{table_fitfunctions_pixel} and~\ref{table_fitfunctions_charge}. For some of the spectra polynomial
terms are added to the fit to describe the background in the spectra most
probably caused by remnant double events. 

\begin{table}
\begin{center}
\begin{tabular}{c|c}
setup&fit function\\
\hline
A&$EG^\text{Cu,esc}_{\text{K}_\alpha}(a,b,N,\mu,\sigma) + EG^\text{Cu}_{\text{K}_\alpha}(a,b,N,\mu,\sigma)$\\
\hline
B&$EG^\text{Mn,esc}_{\text{K}_\alpha}(a,b,N,\mu,\sigma) + EG^\text{Mn}_{\text{K}_\alpha}(a,b,N,\mu,\sigma)$\\
\hline
C&$G^\text{Ti,esc}_{\text{K}_\alpha}(N,\mu,\sigma)+G^\text{Ti,esc}_{\text{K}_\beta}(N,\mu,\sigma)$\\
&$+EG^\text{Ti}_{\text{K}_\alpha}(a,b,N,\mu,\sigma)+G^\text{Ti}_{\text{K}_\beta}(N,\mu,\sigma)$\\
\hline
D&$EG^\text{Ag}_{\text{L}_\alpha}(a,b,N,\mu,\sigma)+G^\text{Ag}_{\text{L}_\beta}(N,\mu,\sigma)$\\
\hline
E&$EG^\text{Al}_{\text{K}_\alpha}(a,b,N,\mu,\sigma)$\\
\hline
F&$G^\text{Cu}_{\text{L}_{\alpha,\beta}}(N,\mu,\sigma)$\\
\hline
G&$G^\text{O}_{\text{K}_\alpha}(N,\mu,\sigma)+G^\text{C}_{\text{K}_\alpha}(N,\mu,\sigma)$\\
&$+G^\text{Fe}_{\text{L}_{\alpha,\beta}}(N,\mu,\sigma)+G^\text{Ni}_{\text{L}_{\alpha,\beta}}(N,\mu,\sigma)$\\
\hline
H&$G^\text{C}_{\text{K}_\alpha}(N,\mu,\sigma)+G^\text{O}_{\text{K}_\alpha}(N,\mu,\sigma)$
\end{tabular}
\end{center}
\caption{Fit functions used for the pixel spectra in Figs.~\ref{fig_spectra1} and~\ref{fig_spectra2}. A Gaussian with amplitude $N$, mean $\mu$ and width $\sigma$ is abbreviated with $G(N,\mu,\sigma)$ while the Gaussian joined with an exponential decay to the left (see Equation~\ref{eq_expogauss}) is noted as $EG(a,b,N,\mu,\sigma)$. The upper and lower indices of the parameters are noted at the function itself as all parameters of a function share the same indices, e.g. $G(N_\alpha,\mu_\alpha,\sigma_\alpha)$ will be noted as $G_\alpha(N,\mu,\sigma)$. Not all parameters were left free for the fits, Table~\ref{table_Xray_lines} lists the parameters fixed for each setting.}
\label{table_fitfunctions_pixel}
\end{table}

\begin{table}
\begin{center}
\begin{tabular}{c|c}
setup&fit function\\
\hline
A&$G^\text{Cu,esc}_{\text{K}_\alpha}(N,\mu,\sigma) + G^\text{Cu}_{\text{K}_\alpha}(N,\mu,\sigma)$\\
\hline
B&$G^\text{Mn,esc}_{\text{K}_\alpha}(N,\mu,\sigma) + G^\text{Mn}_{\text{K}_\alpha}(N,\mu,\sigma)+p_0+p_1\cdot x+p_2\cdot x^2$\\
\hline
C&$G^\text{Ti,esc}_{\text{K}_\alpha}(N,\mu,\sigma)+G^\text{Ti,esc}_{\text{K}_\beta}(N,\mu,\sigma)$\\
&$+G^\text{Ti}_{\text{K}_\alpha}(N,\mu,\sigma)+G^\text{Ti}_{\text{K}_\beta}(N,\mu,\sigma)$\\
\hline
D&$G^\text{Ag}_{\text{L}_\alpha}(N,\mu,\sigma)+G^\text{Ag}_{\text{L}_\beta}(N,\mu,\sigma)+p_0+p_1\cdot x+p_2\cdot x^2$\\
\hline
E&$G^\text{Al}_{\text{K}_\alpha}(N,\mu,\sigma)+p_0+p_1\cdot x+p_2\cdot x^2$\\
\hline
F&$G^\text{Cu}_{\text{L}_{\alpha,\beta}}(N,\mu,\sigma)$\\
\hline
G&$G^\text{O}_{\text{K}_\alpha}(N,\mu,\sigma)$\\
\hline
H&$G^\text{C}_{\text{K}_\alpha}(N,\mu,\sigma)+G^\text{O}_{\text{K}_\alpha}(N,\mu,\sigma)$
\end{tabular}
\end{center}
\caption{Fit functions used for the charge spectra on Figs.~\ref{fig_spectra1} and~\ref{fig_spectra2}. A Gaussian with amplitude $N$, mean $\mu$ and width $\sigma$ is abbreviated with $G(N,\mu,\sigma)$. The upper and lower indices of the parameters are noted at the function itself as all parameters of a function share the same indices, e.g. $G(N_\alpha,\mu_\alpha,\sigma_\alpha)$ will be noted as $G_\alpha(N,\mu,\sigma)$. Not all parameters were left free for the fits, Table~\ref{table_Xray_lines} lists the parameters fixed for each setting.}
\label{table_fitfunctions_charge}
\end{table}

In case of the pixel spectra, the lines of higher X-ray energies show a tail
towards lower energies. This can be explained by insufficient diffusion, which
leads to two electrons being guided into the same grid hole and detected as a
single primary electron. This is the case mostly for more energetic X-rays having
a higher charge density in the center of the charge cloud and a longer
absorption length, resulting in less diffusion. To take this into account, the
Gaussian function was joined by an exponential function: 
\begin{equation}
\label{eq_expogauss}
 f(x)=\left\{\begin{array}{*{2}{c}}
N\exp^{-\frac{(x-\mu)^2}{2\sigma^2}}&\mathrm{for:}~x>c\\\exp^{ax+b}&\mathrm{for:}~x<c\\\end{array}\right.
\end{equation}
$a$ and $b$ parameterize the exponential decay. The parameter $c$, defining
the junction point of the two functions, is determined by the parameters of
the two functions to allow for a continuous function. 

\begin{figure}
\centering
\subfloat[]{\label{fig_calibration_Q}\includegraphics[width=.8\textwidth]{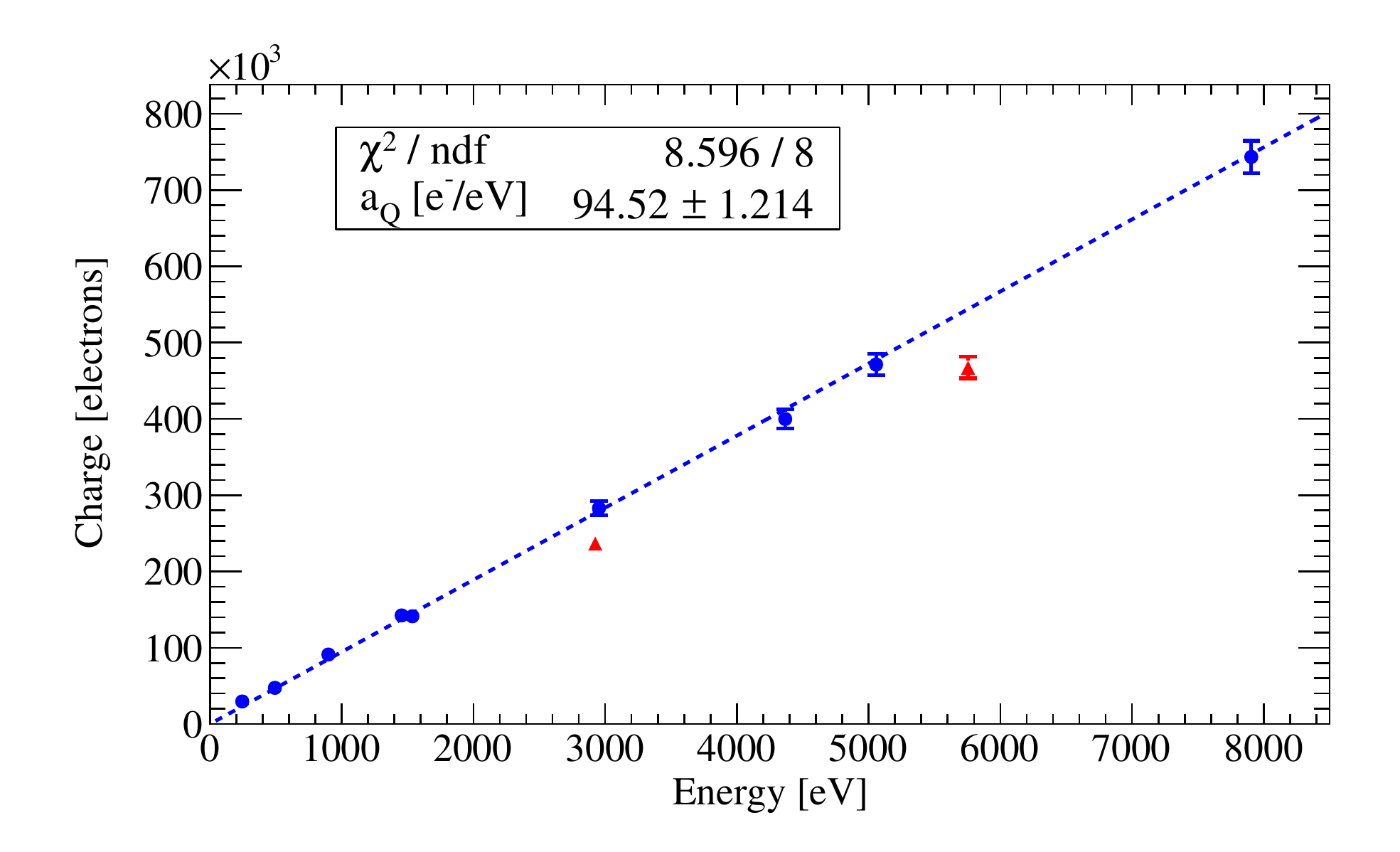}}\\
\subfloat[]{\label{fig_calibration_N}\includegraphics[width=.8\textwidth]{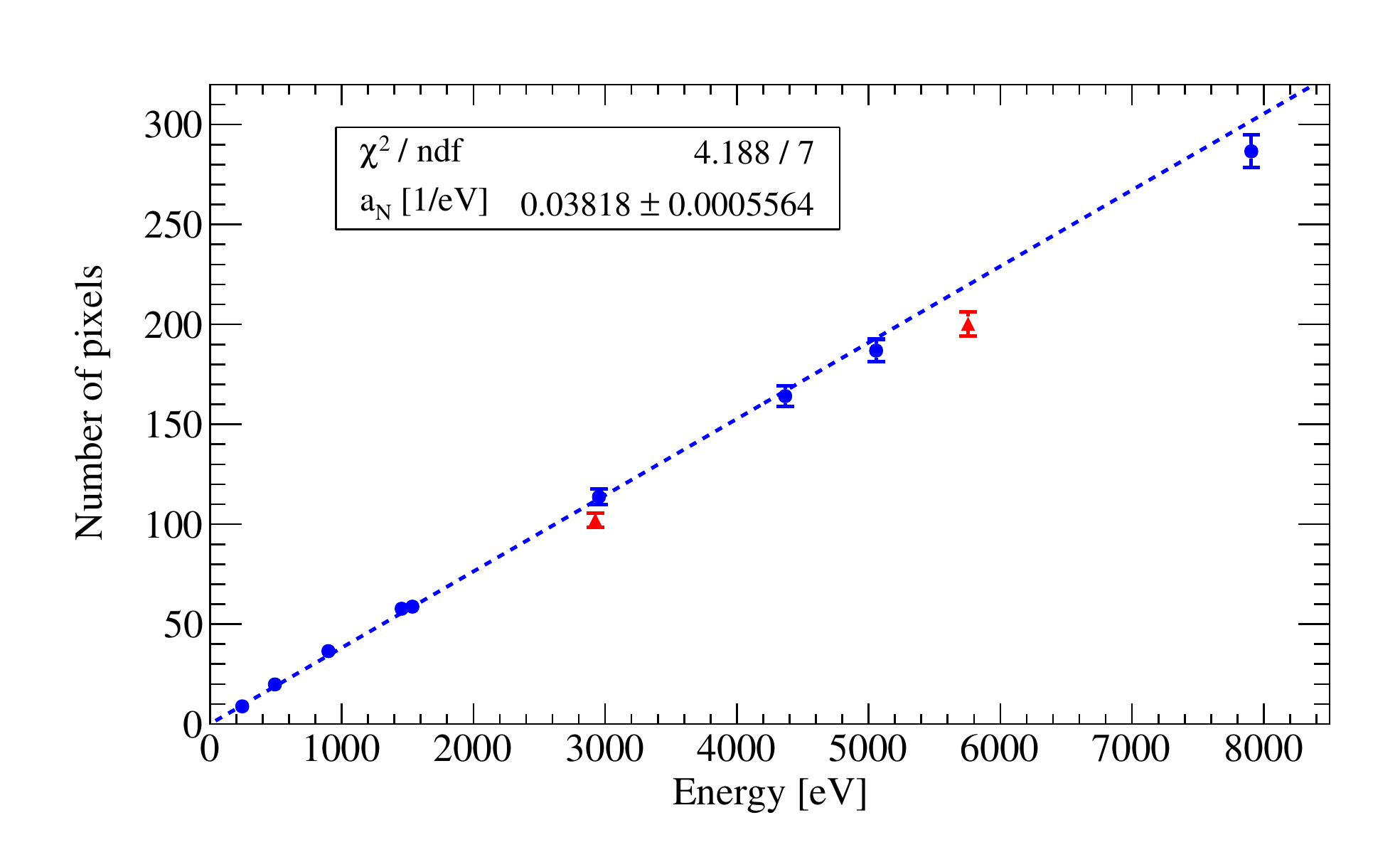}}
\caption{Graphs showing the total charge
  \protect\subref{fig_calibration_Q} and number of activated pixels \protect\subref{fig_calibration_N} in dependence on the X-ray photon
  energy. The manganese lines are not used for the fits and marked as red triangles. The point corresponding to the \SI{8}{\keV} copper line in \protect\subref{fig_calibration_N} was excluded from the fit.}
\label{fig_calibration}
\end{figure}

The fitted peak positions of different X-ray photon energies are shown in
Fig.~\ref{fig_calibration} resulting in a calibration curve for
both the total charge and the number of activated pixels.

The $y$-intercept was set to zero and only the slope $a$ was determined by the
fit giving
the conversion factor form either charge ($a_Q$) or number of pixels ($a_N$) to X-ray photon
energy. In both graphs the measurement with the manganese K$_\alpha$ line at \SI{5.9}{\keV} is
deviating from the fit curve. It was found that the temperature in the laboratory, and consequently in the detector as well, was lower by approximately \SIrange[range-units = single]{5}{10}{\celsius} during this measurement and therefore the gas gain was reduced. It can also be observed, that for high energies (e.g. \SI{8}{\keV}) the number of
pixels is below the calibration curve, because for large enough numbers of primary electrons the diffusion is always insufficient to prevent multiple electrons entering the same grid hole in the center of the photon clusters. To
avoid any bias in the calibration curve, all manganese lines as well as the \SI{8}{\keV} in
the pixel spectrum were not considered in the fit. 

Since the slopes of the calibration curves give the number of pixels per \SI{}{\eV} and the measured charge per \SI{}{\eV} respectively, the gas gain can be determined by $G_a=a_Q/a_N\approx\num{2500}$ from $a_Q$ and $a_N$. The inverse of $a_N$ should also give the mean ionization energy $W_I$ and indeed $a_N^{-1}\approx\SI{26}{\eV}$ which matches the tabulated values for argon and isobutane.

\begin{figure}
\centering
\includegraphics[width=.90\textwidth]{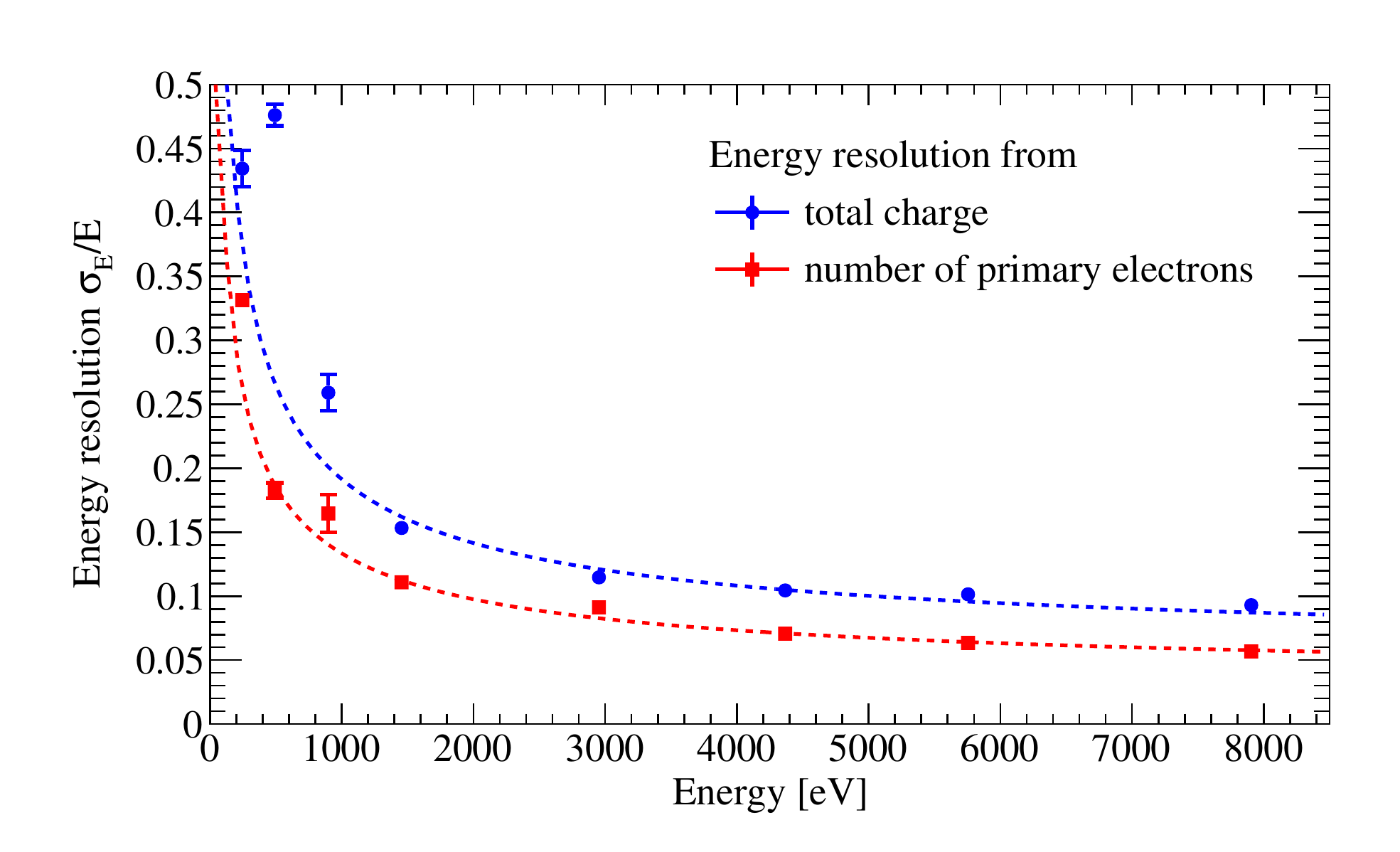}
\caption{Energy resolution $\sigma_E/E$ in dependence of the X-ray photon energy. As the energy can be determined from either the number of activated pixels (detected primary electrons) or the total charge of a reconstructed X-ray photon, the energy resolution for both methods is shown. In both cases the energy dependence is well described by $1/\sqrt{E}$. The deviation of a few points at low energies is caused by those spectra containing additional peaks (e.g. from target contaminations) which cannot be separated from the main peak thus leading to an overestimation of the peak width.}
\label{fig_Eresolution}
\end{figure}

The energy resolution can be defined as $\sigma_E/E = \sigma/\mu$,
where $\sigma$ is the width and $\mu$ the position of the line
under consideration. The values have been extracted from the fits and are shown
for both energy measurement methods in Fig.~\ref{fig_Eresolution}.

The energy resolution improves with $\sqrt{a^2/E+b^2}$ where $a/\sqrt{E}$ describes the statistical part of the energy resolution. As expected the resolution improves with the number of primary electrons ($E\propto N$). $b$ denotes the systematical contribution to the total energy resolution which is added quadratically.

\begin{figure}
\centering
\includegraphics[width=.90\textwidth]{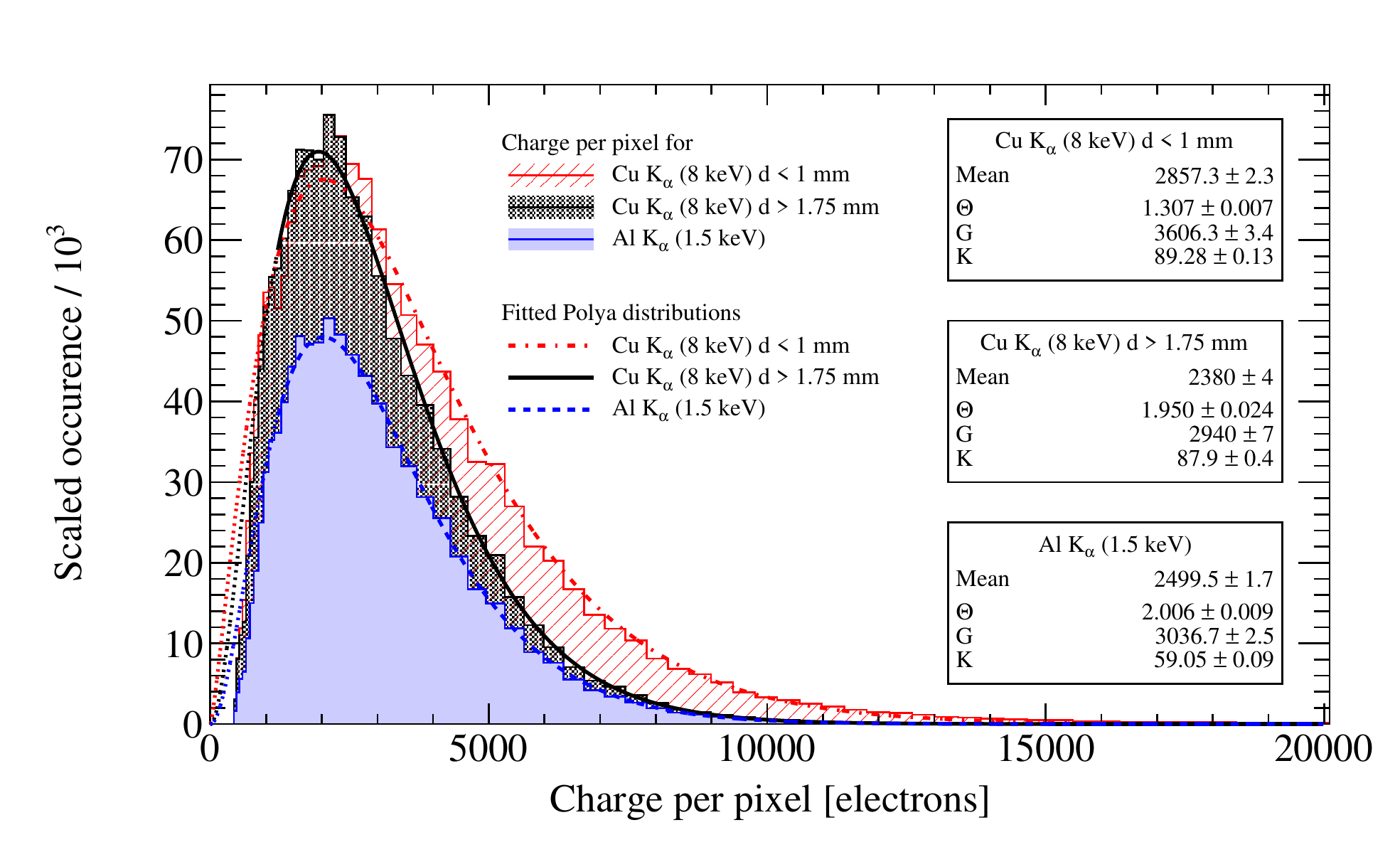}
\caption{Charge distribution collected by individual pixels in events with \SI{1.5}{\keV} X-ray photons (solid light blue) and
\SI{8}{\keV} X-ray photons, the latter split into two distributions: with all
pixels in more than \SI{1.75}{\milli\metre} distance to the reconstructed
center (black dots) and those closer than \SI{1}{\milli\metre} to the center
(red hashes). For each histogram a Polya distribution was fitted to the data,
the resulting parameters are given in the boxes on the right.} 
\label{fig_polya}
\end{figure}

Finally, the gas amplification can also be extracted independently of the
calibration curve by histogramming the charge collected by each
pixel. Fig.~\ref{fig_polya} shows three different distributions: one filled with pixels' charges from \SI{1.5}{\keV} X-ray events, two filled with charges collected on individual pixels from \SI{8}{\keV} X-ray events but only using those primary electrons (pixels) which are at
least \SI{1.75}{\milli\metre} from the event center and those with a maximum distance of \SI{1}{\milli\metre} to the event center respectively.
The distributions are fitted to a Polya-function. The
parameterization is given by~\cite{blumrolandi}:
\begin{equation}
P_\text{Polya}(x)=\frac{K}{G_\text{Polya}}\frac{(\Theta+1)^{\Theta+1}}{\Gamma(\Theta+1)}\left(\frac{x}{G_\text{Polya}}\right)^\Theta\exp\left(-(\Theta+1)\frac{x}{G_\text{Polya}}\right)
\end{equation}
where $K$ is a scaling parameter to adopt the normalized distribution to the
data, $G_\text{Polya}$ is a measure for the gas gain (but cannot be directly compared to the mean or MPV of the distribution) and $\Theta$ is inverse
proportional to the width of the distribution. As expected the first two
distributions are in good agreement and show a Polya distribution. From the
fit the gas gain can be extracted to be
$G_\text{Polya}\approx\num{3000}$. Taking the mean of the distributions one gets $G_\text{mean}\approx\num{2500}$ ($G_\text{MPV}\approx\num{2200}$) which is in
good agreement with $G_a$. However, the distribution given by pixels in the
center of the \SI{8}{\keV} X-ray photon charge clouds becomes notably
broader. This can be explained by contributions of pixels with two electrons
collected. This also leads to an overestimation of the mean gas gain.

\section{Additional Features of the Events}\label{sec_addFeatures}
\begin{figure}
\centering
\subfloat[]{\label{fig_diffusion_al}\includegraphics[width=.48\textwidth]{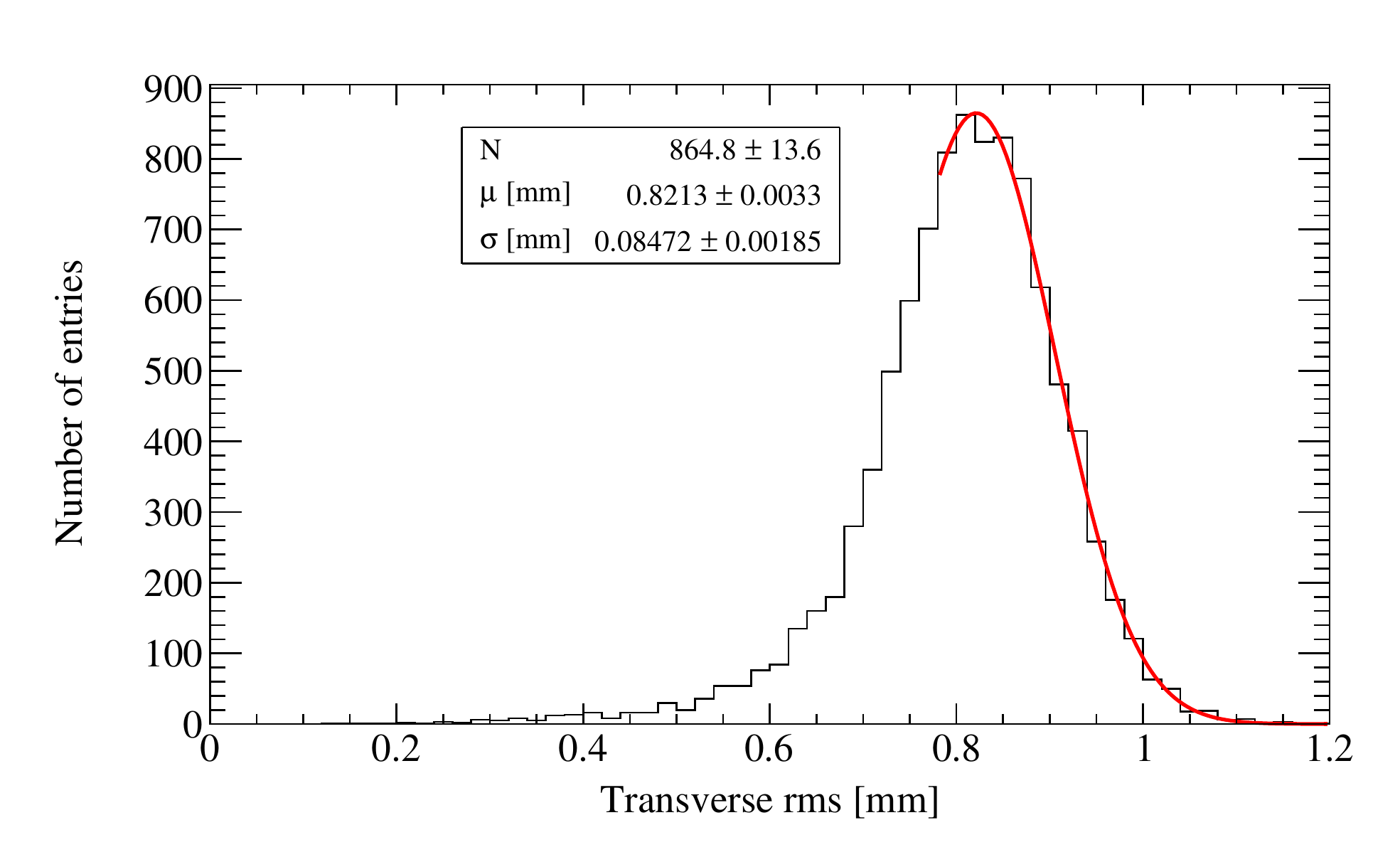}}
\subfloat[]{\label{fig_diffusion_cu}\includegraphics[width=.48\textwidth]{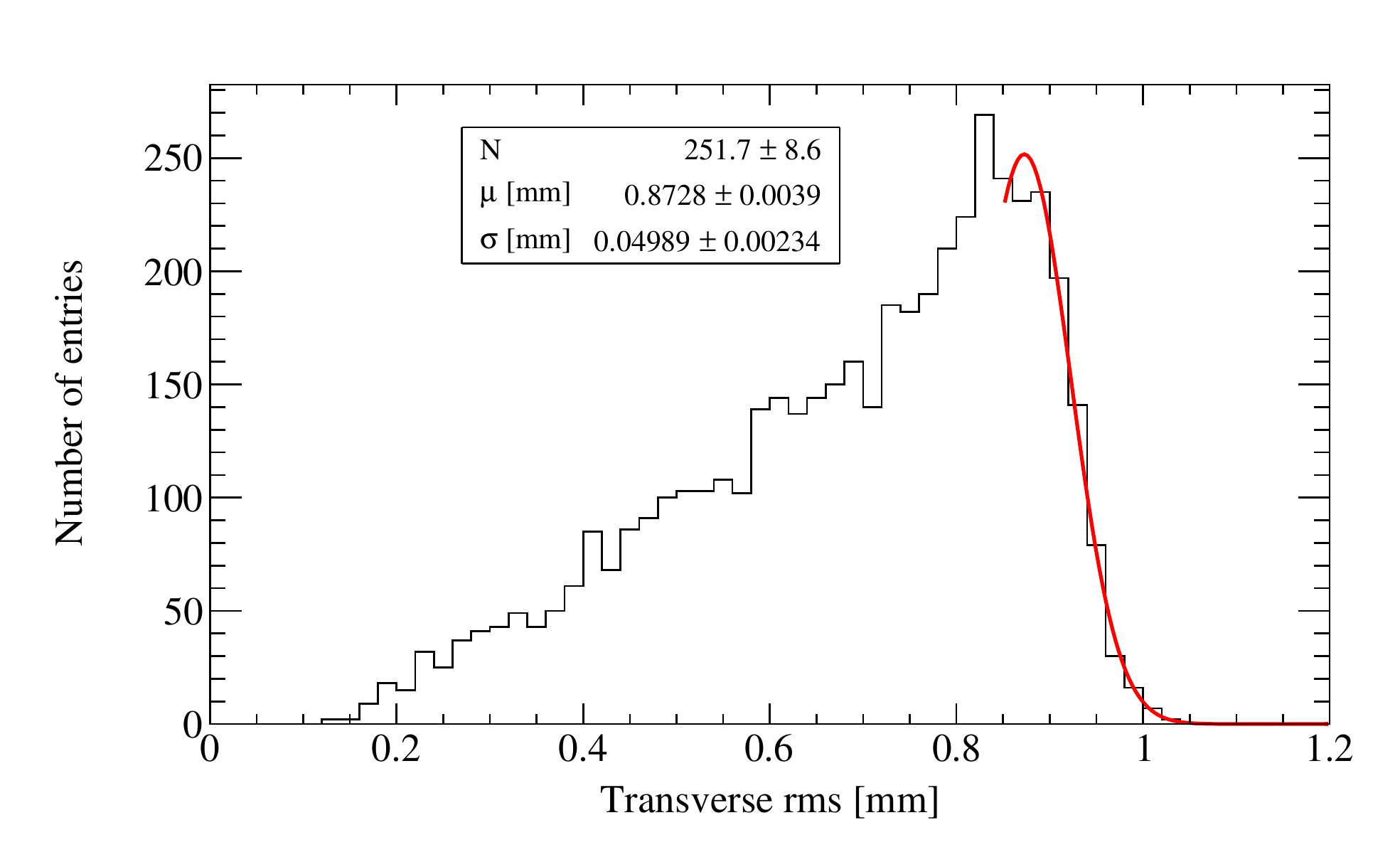}}
\caption{Diffusion of the charge cloud as defined by $\sigma_\text{trans}$. Distribution for \SI{1.5}{\keV} photons \protect\subref{fig_diffusion_al} and
for \SI{8}{\keV} photons \protect\subref{fig_diffusion_cu} are shown. The fit to the right side gives the maximum diffusion.}
\label{fig_diffusion}
\end{figure}

 Fig.~\ref{fig_diffusion} shows the histogram of
the cluster widths $\sigma_\text{trans}$ for two different X-ray energies. 
 Both distributions reach a maximum slightly above \SI{0.8}{\milli\metre} which is the maximal
 cluster size after a \SI{3}{\centi\metre} drift (calculation of $\sigma(\SI{3}{\centi\metre})$ see
 Sect.~\ref{sub_sec_analysis}). A Gaussian fit to the right side of the
 distribution determines the maximum of the distribution and thus gives the
 diffusion coefficient. The fits result in $D_t\approx\SI[per-mode=fraction]{474}{\micro\metre\per\sqrt{\centi\metre}}$ and $D_t\approx\SI[per-mode=fraction]{504}{\micro\metre\per\sqrt{\centi\metre}}$ respectively. The slight variations can be attributed to temperature changes between the different measurements.
 
 For the low energetic X-rays a fast drop on
 the left side indicates, that most of the photons are absorbed close to the
 cathode, while the slower drop for higher energetic photons shows the higher
 penetration power of these photons and thus the reduced diffusion for these
 electron clouds.  

\begin{figure}
\centering
\subfloat[]{\label{fig_logl_fraction_al}\includegraphics[width=.48\textwidth]{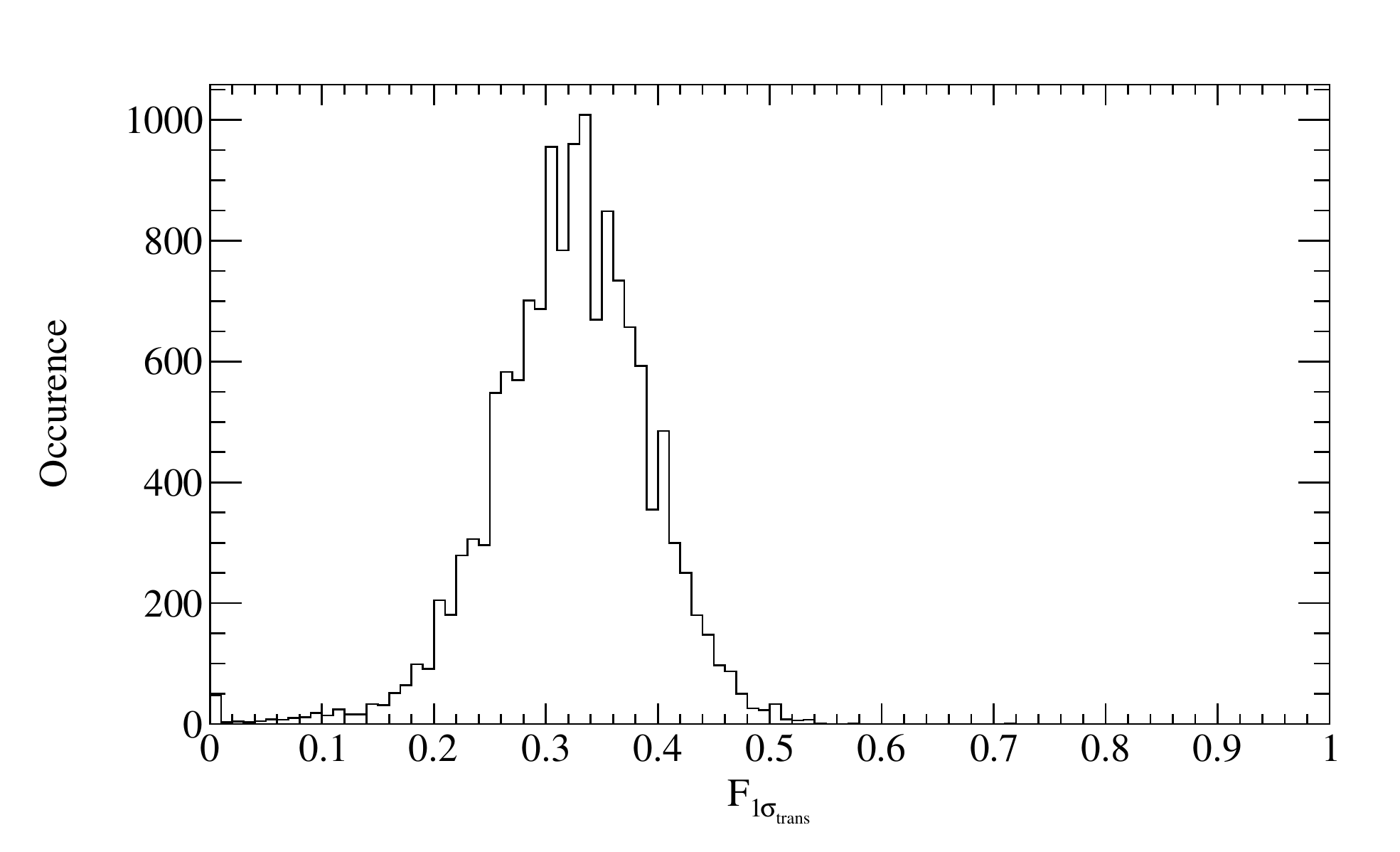}}
\subfloat[]{\label{fig_logl_fraction_mean_rms}\includegraphics[width=.48\textwidth]{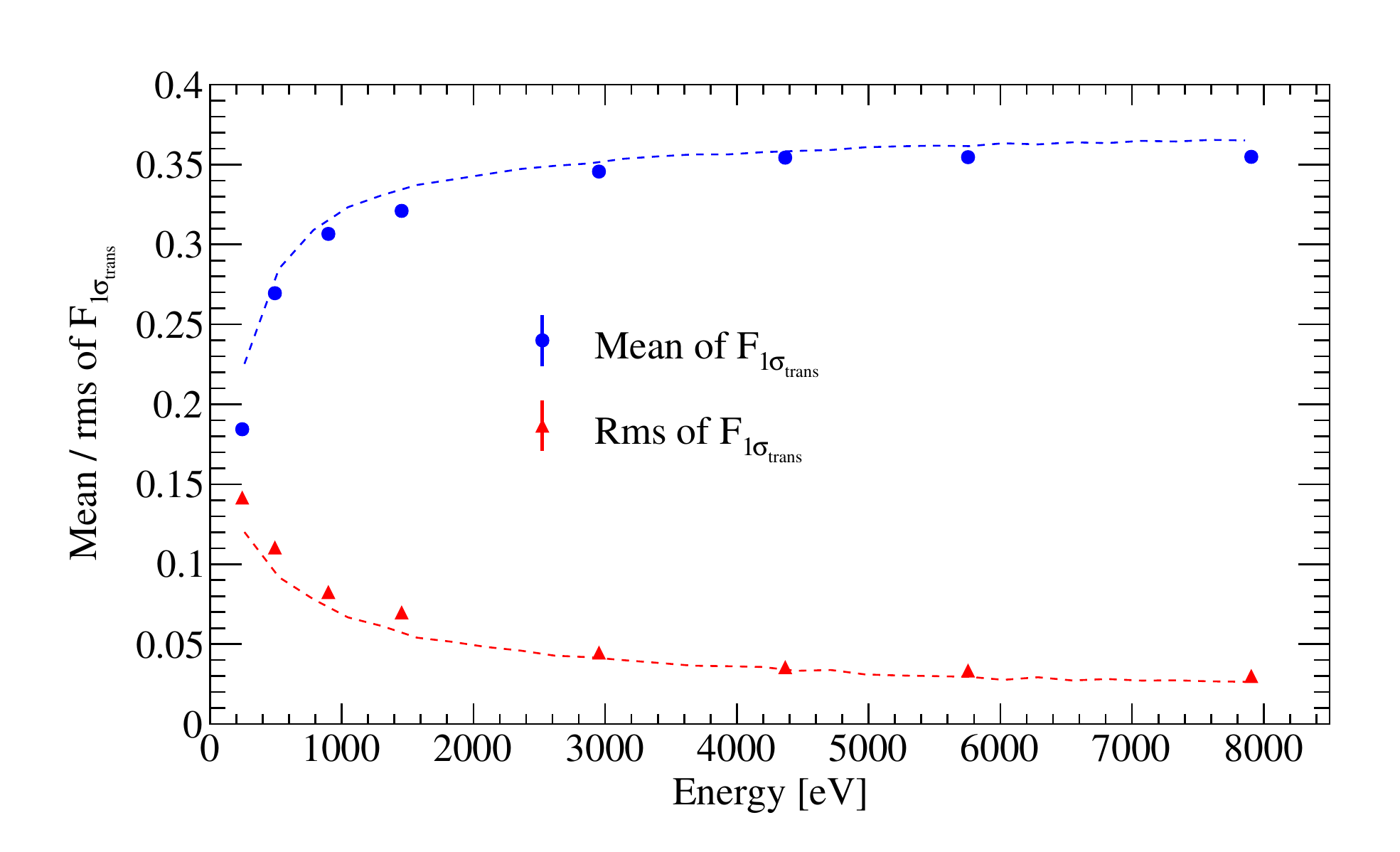}}\\
\subfloat[]{\label{fig_logl_exc_al}\includegraphics[width=.48\textwidth]{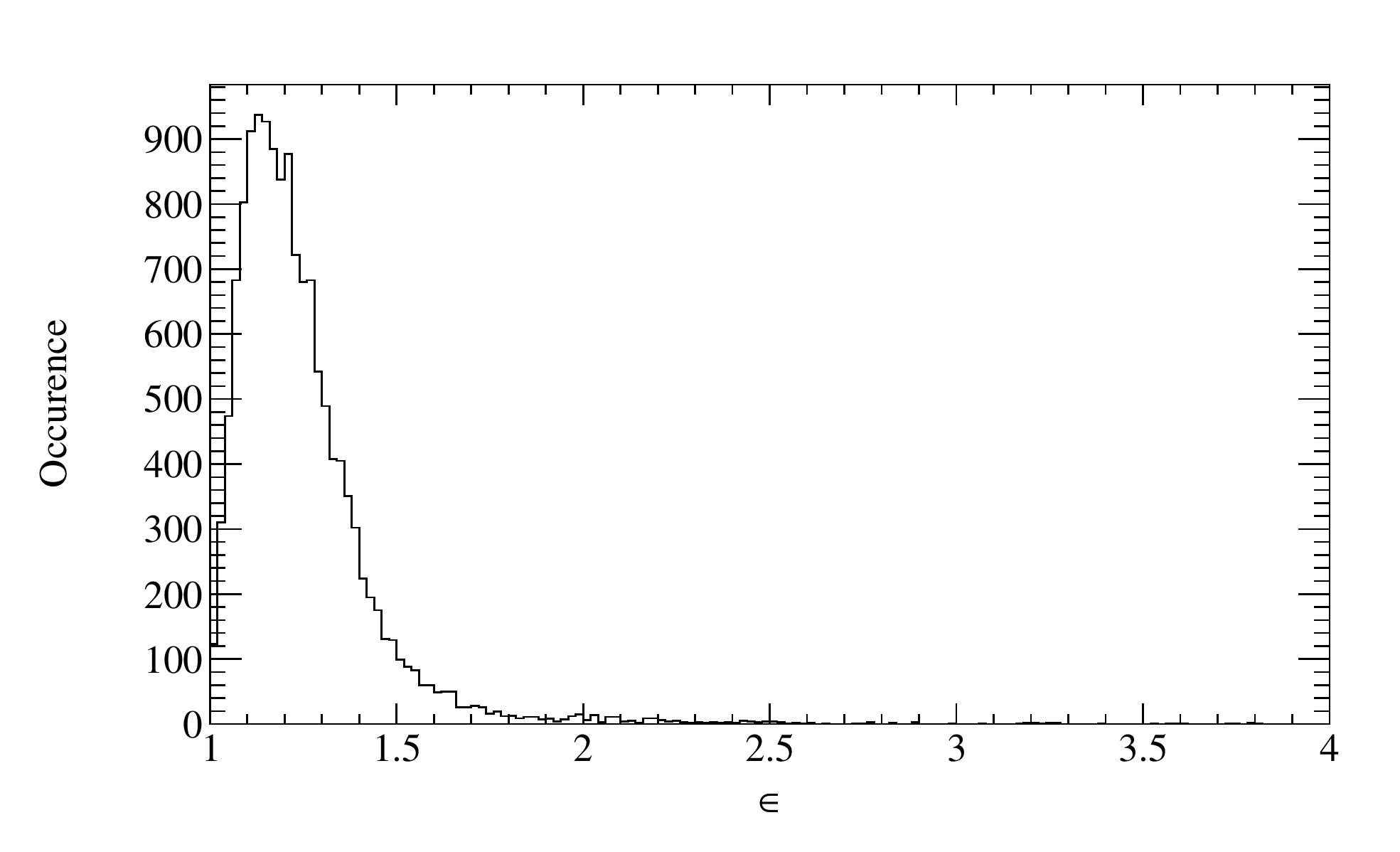}}
\subfloat[]{\label{fig_logl_exc_mean_rms}\includegraphics[width=.48\textwidth]{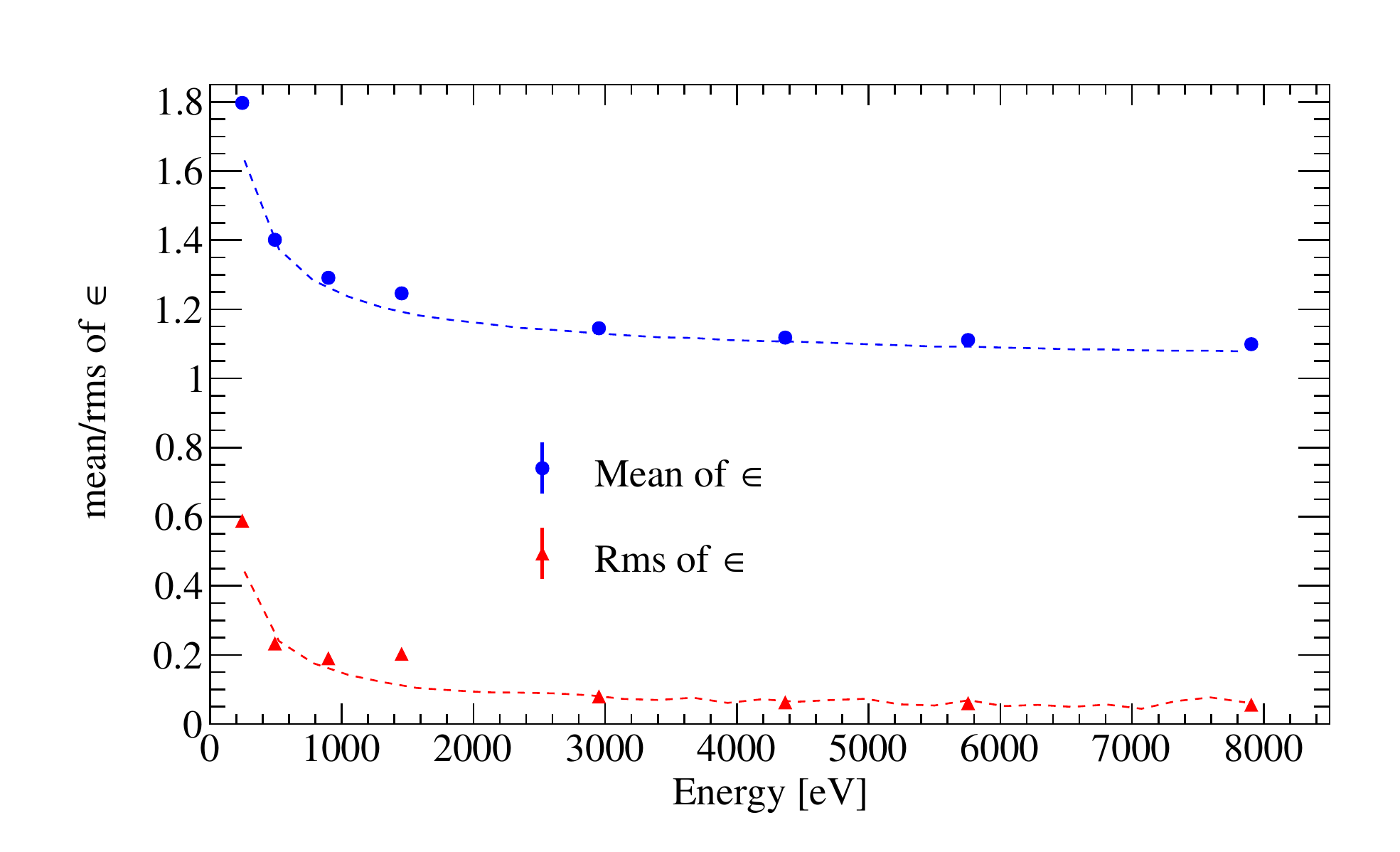}}\\
\subfloat[]{\label{fig_logl_length_al}\includegraphics[width=.48\textwidth]{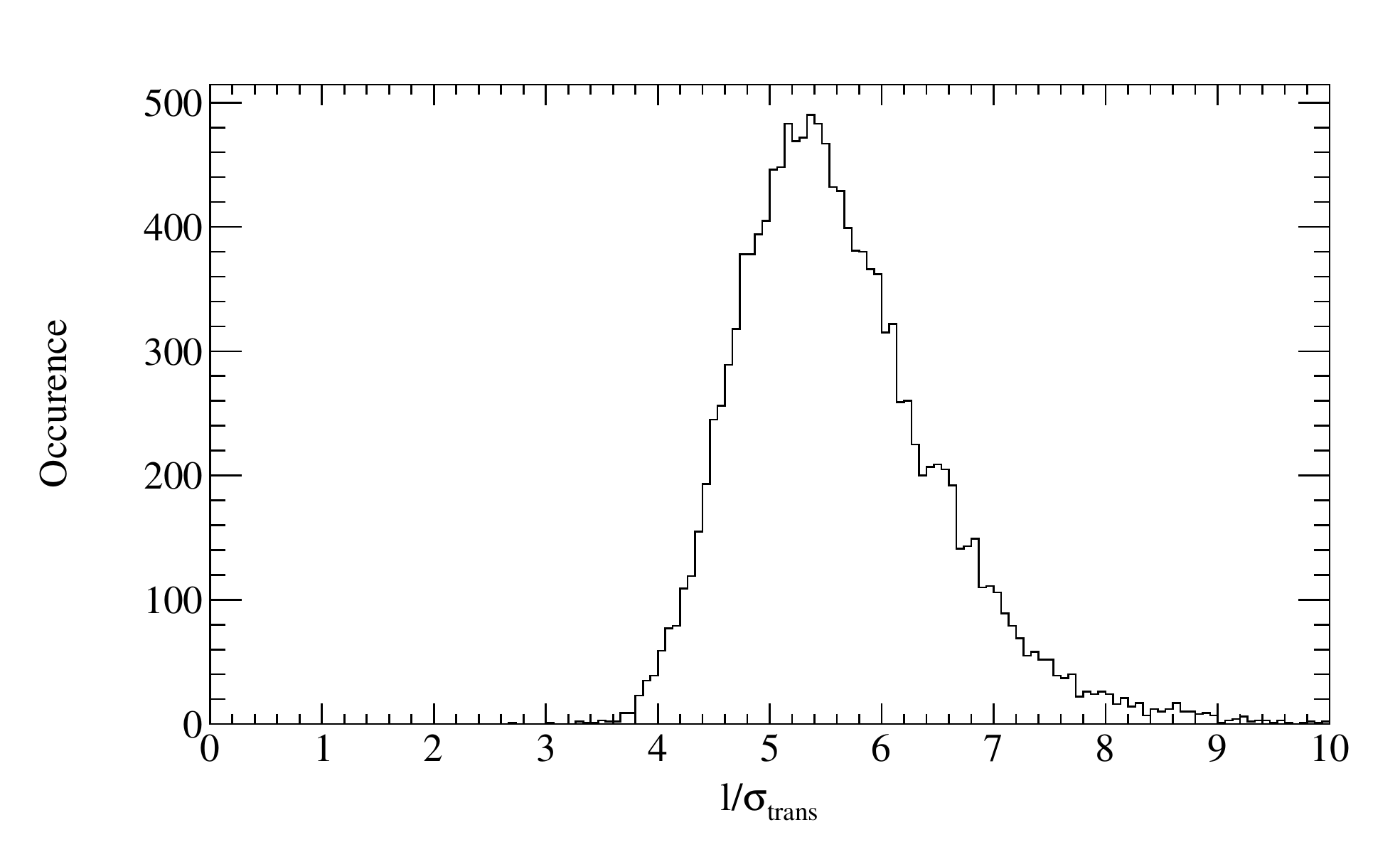}}
\subfloat[]{\label{fig_logl_length_mean_rms}\includegraphics[width=.48\textwidth]{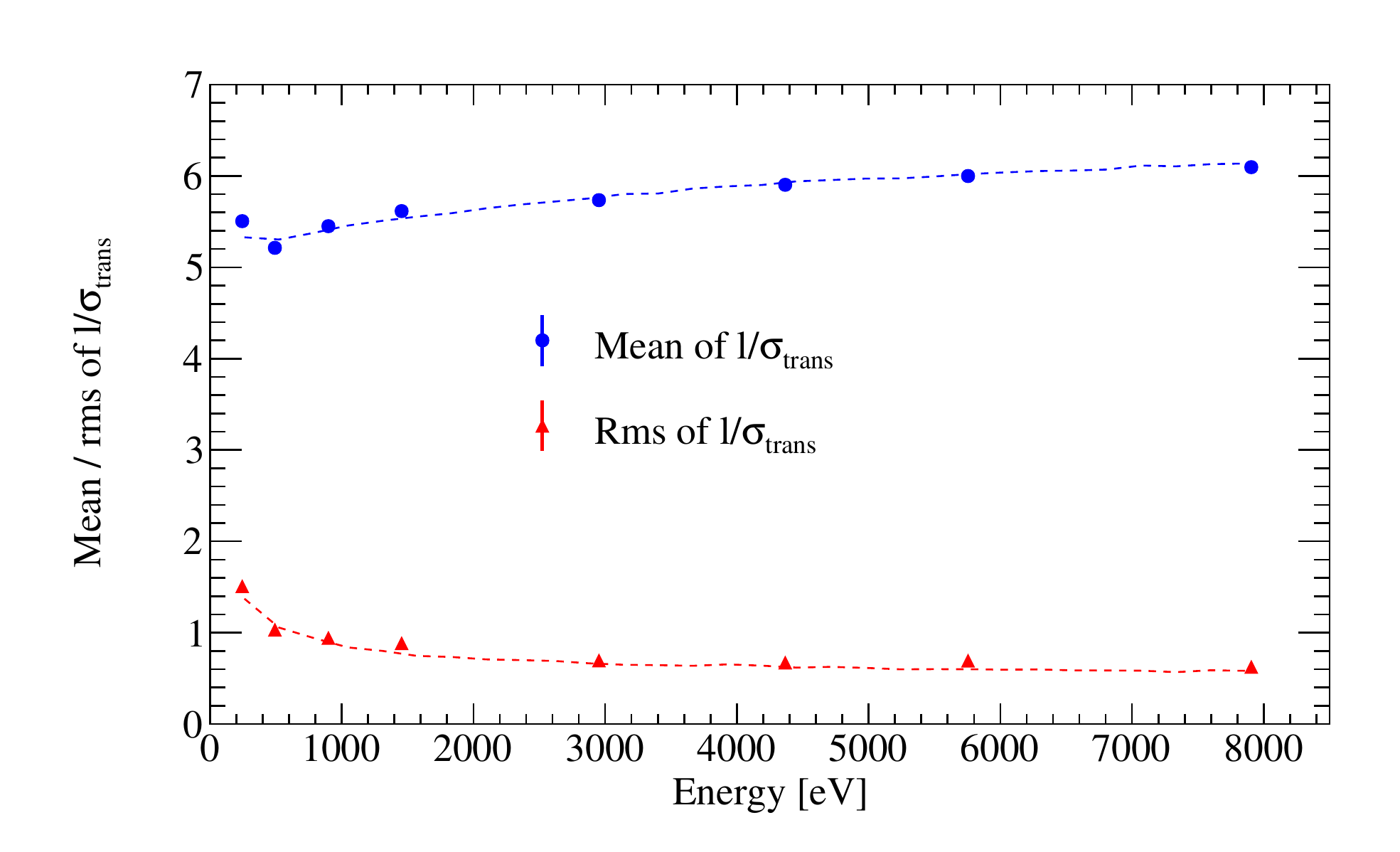}}
\caption{Energy dependence of three eventshape variables. On the left side are example
  distributions of the \SI{1.5}{\keV} X-rays, on the right side the energy dependencies of
the mean and the width are given. The thin dashed lines show the results of a
simple Monte Carlo simulation described in the text. The three variables are: fraction $F_{\num{1}\sigma_\text{trans}}$ of pixels within \num{1} rms around the reconstructed center, \protect\subref{fig_logl_fraction_al} and \protect\subref{fig_logl_fraction_mean_rms}), eccentricity $\epsilon$, \protect\subref{fig_logl_exc_al} and \protect\subref{fig_logl_exc_mean_rms}, and length $l$ divided by transverse rms $\sigma_\text{trans}$, \protect\subref{fig_logl_length_al} and \protect\subref{fig_logl_length_mean_rms}.}
\label{fig_logl1}
\end{figure}

The energy dependence of the three eventshape parameters introduced in section~\ref{sub_sec_analysis} are shown in Fig.~\ref{fig_logl1}. The left
column always shows distribution of the \SI{1.5}{\keV} photons and in the right column
the energy dependence of the means and the width are shown.

For the fraction $F_{\num{1}\sigma_\text{trans}}$ of pixels within \num{1} rms around the center of the charge
cloud, the mean approaches about \SI{35}{\percent} for high energetic photons, which is shown in Fig.~\ref{fig_logl_fraction_mean_rms}. The
eccentricity $\epsilon$ of higher energetic photons is close to one (see Fig.~\ref{fig_logl_exc_mean_rms}) indicating a good
circularity. The length $l$ divided by $\sigma_\text{trans}$ varies between \num{5} and
\num{6} (see Fig.~\ref{fig_logl_length_mean_rms}), which shows that all electrons are contained in the range of $\pm \num{3}~\sigma_\text{trans}$ as expected from statistical considerations. For tracks
parallel to the grid the expected numbers are significantly 
smaller for $F_{\num{1}\sigma_\text{trans}}$, larger (up to about
\num{12}) for $\epsilon$ and also for $l/\sigma_\text{trans}$ (up to approximately \num{35} to \num{40}).

For all three variables, the separation power decreases for lower energetic
photons, since the widths of the distributions increases and also the central
value shifts. This is due to the smaller number of the primary electrons,
where individual electrons experiencing a higher diffusion have more impact.

A simple Monte Carlo simulation was used to study the dependence of the
quantities on the statistics. A fixed number of electrons increasing in steps
of \num{10} from \num{10} to \num{300} was smeared by a 2D Gaussian distribution with a width
corresponding to the maximum diffusion $\sigma(\SI{3}{\centi\metre})\approx\SI{800}{\micro\metre}$. The
final position of each electron was quantized in steps of
$\SI{55}{\micro\metre}$ representing the finite pitch of the GridPix
detector. The reconstruction and analysis was done with the same software as for
the detector data. The values of the simulations are indicated by thin dashed connecting lines in all graphs. They follow the shape of the curves given
by the detector data and small quantitative deviation can be explained by the limited accuracy of the simulation (e.g. neglecting the initial track of the photoelectron in the conversion).
 
\begin{figure}
\centering
\subfloat[]{\label{fig_logl_skew_al}\includegraphics[width=.48\textwidth]{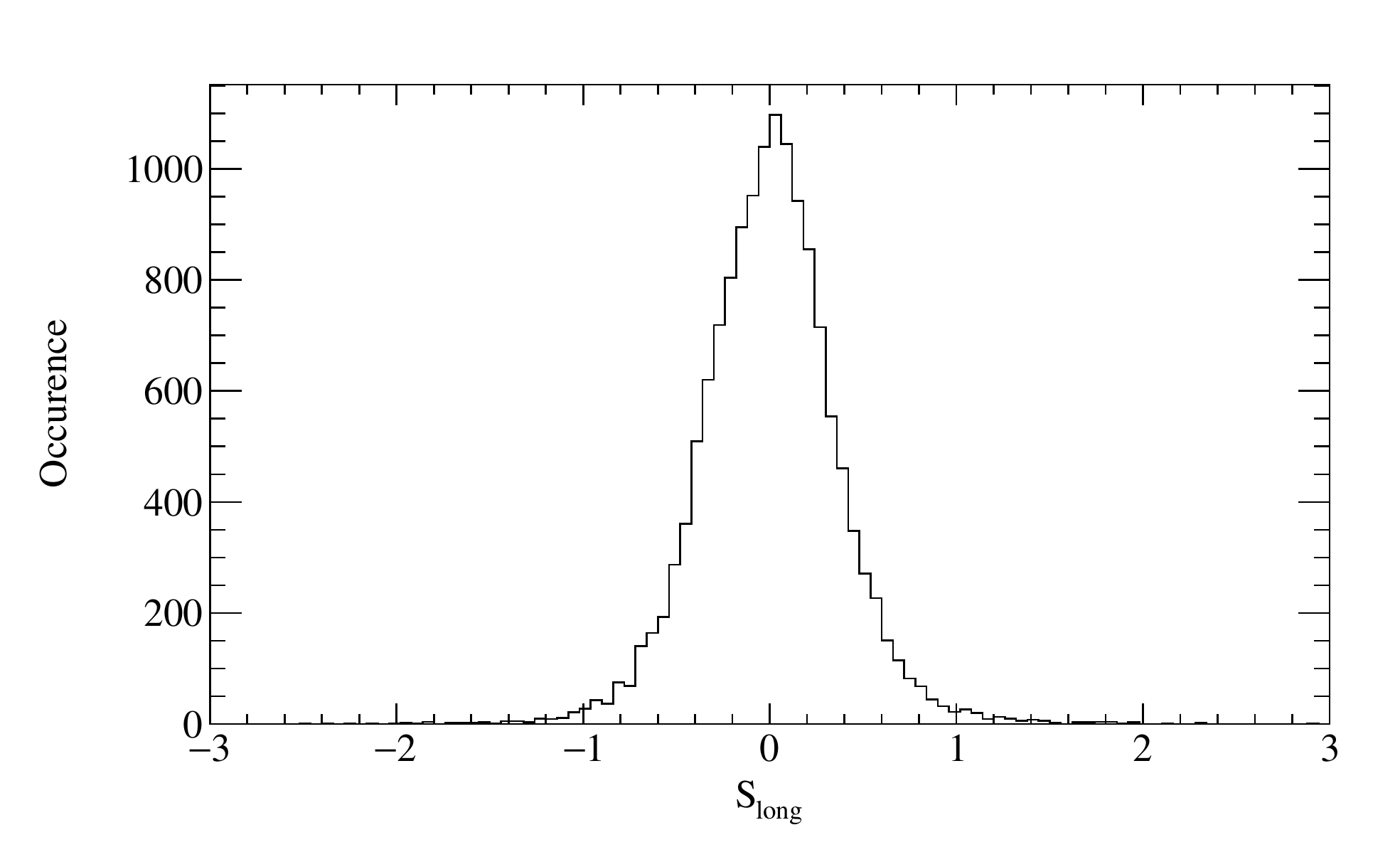}}
\subfloat[]{\label{fig_logl_skew_mean_rms}\includegraphics[width=.48\textwidth]{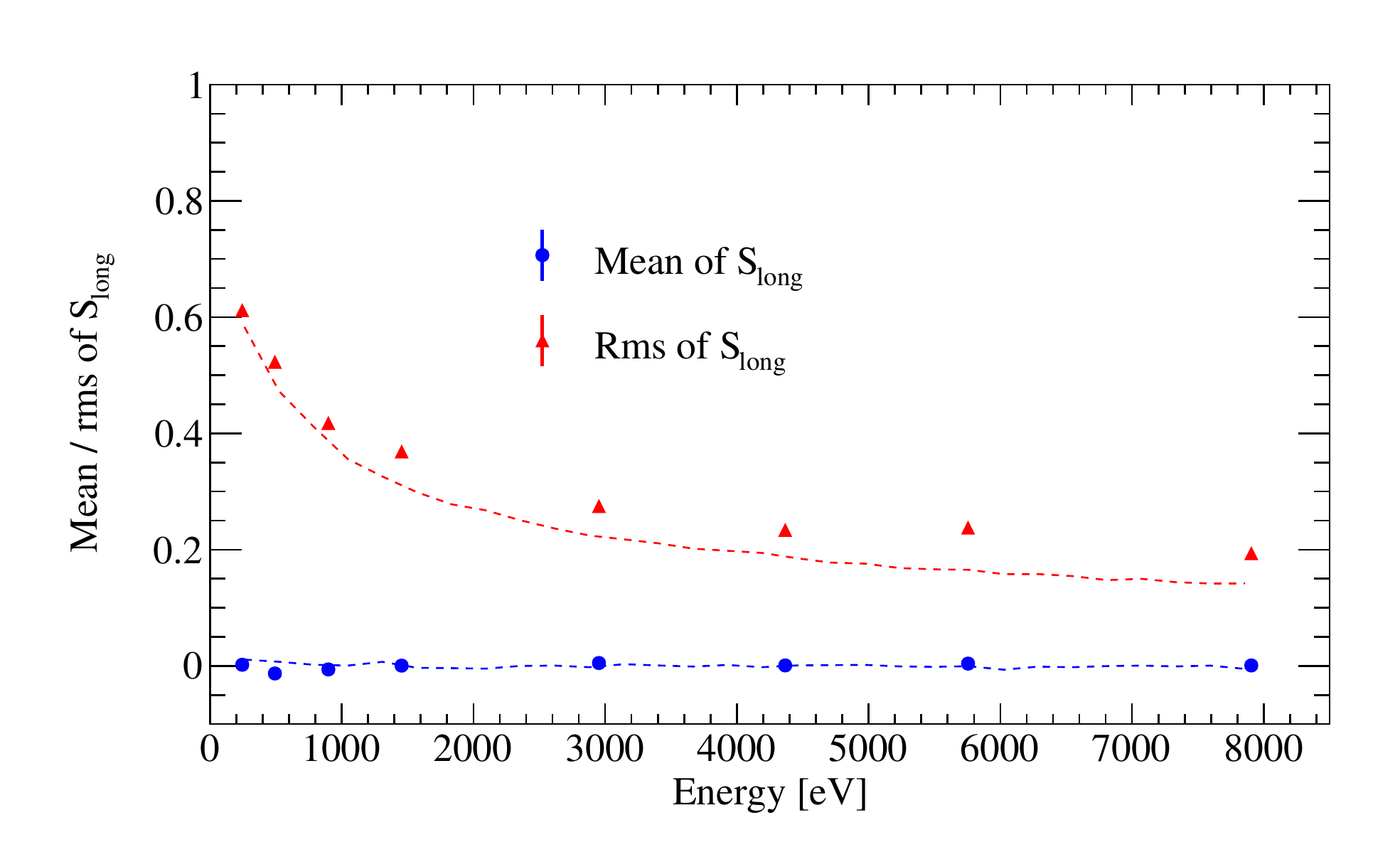}}\\
\subfloat[]{\label{fig_logl_kurt_al}\includegraphics[width=.48\textwidth]{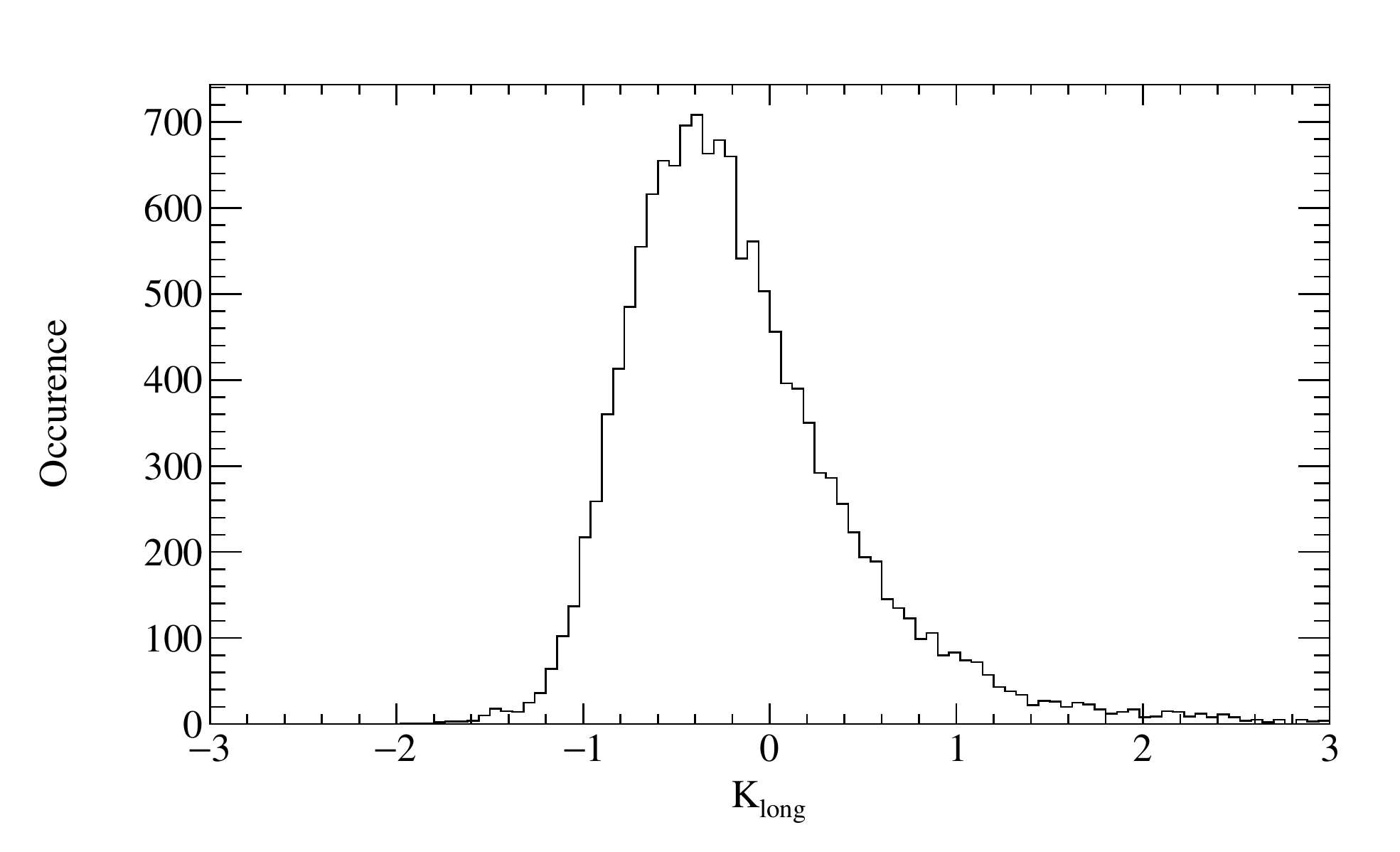}}
\subfloat[]{\label{fig_logl_kurt_mean_rms}\includegraphics[width=.48\textwidth]{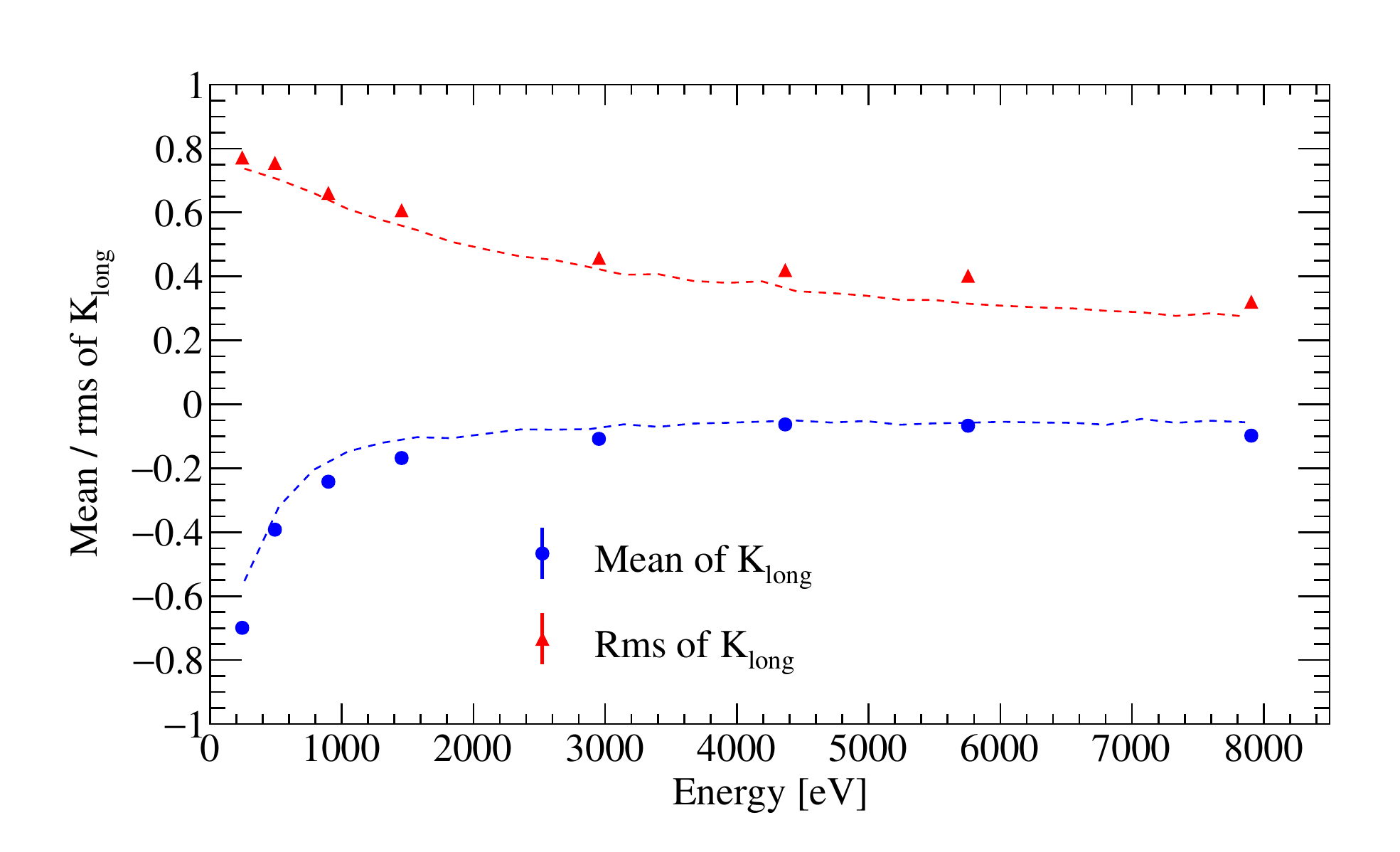}}
\caption{The two central moments skewness $S_\text{long}$, \protect\subref{fig_logl_skew_al} and
  \protect\subref{fig_logl_skew_mean_rms}, and excess kurtosis $K_\text{long}$, \protect\subref{fig_logl_kurt_al} and \protect\subref{fig_logl_kurt_mean_rms}. On the left side are example
  distributions of the \SI{1.5}{\keV} X-rays, on the right side the energy dependencies of
the mean and the width are given. The thin dashed lines show the results of a
simple Monte Carlo simulation described in the text.}
\label{fig_logl2} 
\end{figure}

The third and fourth central moments, skewness $S$ and excess kurtosis $K$, give additional
information about the shape of the distribution. Since they are defined in one
dimension only, the projection of the hits on the long axis has been
chosen. These measures are defined as
\begin{equation}
S_\text{long}=\frac{1}{n}\sum^n_{i=0}\frac{(x_i - \mu_x)^3}{\sigma_x^3}
\end{equation}
and
\begin{equation}
K_\text{long}=\left(\frac{1}{n}\sum^n_{i=0}\frac{(x_i - \mu_x)^4}{\sigma_x^4}\right)-3
\end{equation}
where $x_i$ are the positions of the pixels on the long axis, $\mu_x$ is the mean of the $x_i$ and $\sigma_x=\sigma_\text{long}$ the rms along the long axis.
Fig.~\ref{fig_logl2} shows the two quantities on the left side for the
\SI{1.5}{\keV} aluminum K$_\alpha$ line, while on the right side the dependence on the energy is
shown. 

The skewness is centered around \num{0} indicating that the signals are symmetric
around the reconstructed central position. Except a widening of the
distribution for lower energetic photons, no variations can be observed. Since
tracks are most of the times also symmetric around the reconstructed center,
the quantity can not be used for a CAST background suppression.

In Fig.~\ref{fig_logl2} the excess kurtosis is shown as well. Its mean is around zero for higher
energetic photons as expected for a normal distribution. However, lower
energetic photons show a mean significantly lower, which is a sign of a
flatter top than a normal distribution. This is due to a stronger impact of
single outlying electrons, determining the long axis and stretching the scale
systematically selected for this quantity. 

\section{Conclusion}\label{sec_conc}
An argon/isobutane (\SI{97.7}{\percent}/\SI{2.3}{\percent})-filled GridPix detector has been successfully tested and calibrated with soft X-rays (\SI{277}{\eV} to \SI{8}{\keV}) using the variable energy X-ray source of the CAST Detector Lab at CERN. A good linearity and energy resolution could be demonstrated using both number of hit pixels and total measured charge as energy measures. An energy resolution ($\sigma_E/E$) better than \SI{10}{\percent} (\SI{20}{\percent}) was reached for energies above \SI{2}{\keV} (\SI{0.5}{\keV}). Several characteristics of the X-ray events were studied and three eventshape variables were identified which can be used for discrimination of non-X-ray background events in an application at a low rate experiment such as CAST where a low background rate is crucial for increasing the sensitivity for axions and other particles beyond the Standard Model. The energy dependence of these variables has been studied and is well understood.

\section*{Acknowledgments}
We thank the CAST collaboration for providing access to the CAST Detector Lab facilities, including especially the X-ray generator, and the support in preparing and performing the measurements.

\section*{References}

\bibliography{nima2017xray}

\end{document}